\documentclass[letterpaper]{article} 
\usepackage{aaai23}  
\usepackage{times}  
\usepackage{helvet}  
\usepackage{courier}  
\usepackage[hyphens]{url}  
\usepackage{graphicx} 
\usepackage{natbib}  
\usepackage{caption} 
\usepackage{booktabs}
\usepackage{multirow}
\usepackage{caption}
\usepackage{subcaption}
\usepackage{amsmath}

\usepackage{algorithm}
\usepackage{algorithmic}

\usepackage{newfloat}
\usepackage{listings}
\DeclareCaptionStyle{ruled}{labelfont=normalfont,labelsep=colon,strut=off} 
\floatstyle{ruled}
\newfloat{listing}{tb}{lst}{}
\floatname{listing}{Listing}
\title{The Chance of Winning Election Impacts on Social Media Strategy}
\author {
    Taichi Murayama,\textsuperscript{\rm 1}
    Akira Matsui, \textsuperscript{\rm 2}
    Kunihiro Miyazaki, \textsuperscript{\rm 3} 
    Yasuko Matsubara,\textsuperscript{\rm 1}
    Yasushi Sakurai,\textsuperscript{\rm 1}\\
}
\affiliations {
    \textsuperscript{\rm 1} SANKEN, Osaka University 
    \textsuperscript{\rm 2} Yokohama National University
    \textsuperscript{\rm 3} Indiana University Bloomington \\
    \{taichi,yasuko,yasushi\}@sanken.osaka-u.ac.jp,  matsui-akira-zr@ynu.ac.jp, kunihirom@acm.com
}

\usepackage{bibentry}

\begin{document}

\maketitle

\begin{abstract}
Social media has been a paramount arena for election campaigns for political actors.
While many studies have been paying attention to the political campaigns related to partisanship, politicians also can conduct different campaigns according to their chances of winning. 
Leading candidates, for example, do not behave the same as fringe candidates in their elections, and vice versa. 
We, however, know little about this difference in social media political campaign strategies according to their odds in elections.
We tackle this problem by analyzing candidates' tweets in terms of users, topics, and sentiment of replies.
Our study finds that, as their chances of winning increase, candidates narrow the targets they communicate with, from people in general to the electoral districts and specific persons (verified accounts or accounts with many followers).
Our study brings new insights into the candidates' campaign strategies through the analysis based on the novel perspective of the candidate's electoral situation.
\end{abstract}

\section{Introduction}
Social media has become an important tool in election campaigns.
Political actors receive a variety of benefits from using social media; for example, causing voting behavior~\cite{kovic2020brute}, attracting new party members~\cite{gibson2018friend}, and provoking political debate~\cite{paul2017compass}.
Consequently, they increasingly expect their messages on social media to have the above-mentioned effects.
General users are also often exposed to political topics on social media.
On Twitter, the U.S. election alone was the second most tweeted event in 2016~\cite{independent_news}.
Thus, the way to use social media has a strong influence on politics and election topics.

How electoral candidates handle social media according to their chances of winning an election is unclear.
Existing many studies addressing political communication on social media often focus on binary opposition, such as the two-party system, i.e., ruling and opposition parties~\cite{heiss2019drives,keller2018followers,bobba2019social}.
Not much research has been conducted on fringe candidates' political communication on social media because they have little influence and are unlikely to affect the overall outcome of an election in a two-party system like the U.S.
In recent years, however, they often activate political discussions on social media and turn out their followers, with running as candidates even though they have low chances of winning elections for fusion voting~\cite{fusion} and issue awareness~\cite{issue, kitunzi2016influence}.
The freshness of their slogans and their substandard movement sometimes succeed in attracting people's attention and contributing to their win; for example, in Japan, the Trumpian-inspired party (Sanseito) won a seat in the House of Councilors in 2022~\cite{sanseito}.
Their increasing presence makes it important to understand their behaviors and strategies on social media, which could not be covered by an analysis of the existing two-party system against the backdrop of U.S. society.
Likewise, it is not well understood how leading candidates who rarely lose elections use social media, compared to them.

In this work, we aim to deepen our understanding of the differences in their social media strategies during elections in response to the chances of winning.
To this aim, we collect the candidates' posts and user information in the Japanese Twitter-sphere leading up to the 2022 Upper House election and attempt to characterize them classified into four groups according to the chances of winning (almost winning, even, almost losing, and proportional representation group).


We tackle the following research questions by the comparison between four groups.

\noindent \textbf{RQ1: What are the characteristics of the frequency of tweets and user information?}
We attempt to examine and find a statistical difference between the four groups  according to the chances of winning (almost winning, even, almost losing, and proportional representation group) in basic tweet behavior and user information.
It provides useful insight into how each group is dealing with social media.

\noindent \textbf{RQ2: What kind of content does each group post during the election period?}
We analyze what kind of content is likely to be posted during election periods through the topic model.
By identifying differences in the content that each group is most likely to discuss, it becomes clear what election issues they want the public to pay attention to and what they want to claim.
We expect to see differences in social media strategies based on the chances of winning.

\noindent \textbf{RQ3: What type of content affects user engagement?}
We analyze which content is more likely to gain user engagement through the regression analysis method.
The mechanism of what content they encourage their followers to share their content has so far not been a focus of extensive study.
We seek to fill the gap by understanding how users are likely to respond to content in each group on Twitter communication during an election.
This also helps all parties effectively promote participatory democracy globally and their campaign policies.

\noindent \textbf{RQ4: Is there a difference in the way they communicate with other users on social media?}
One of the most efficient ways for candidates to communicate directly with voters and other candidates is to utilize reply functions.
We attempt to elucidate how the electoral situation makes a difference in the way they communicate with them.

By answering these research questions, we made the following contribution:
\begin{itemize}
    \item We revealed that the number of followers, i.e., the popularity on social media, does not necessarily increase the chances of winning an election.
    Nonetheless, our analysis also finds that candidates with little chance of winning are more aggressive in their social media strategies than other candidates (Section 4).
    \item Candidates in a state of close competition tend to tweet more about the neighborhoods where their constituents live, while candidates with little chance of winning the election tend to post tweets asking all of their followers to vote or share (Section 5).
    \item Tweets asking all of their followers to vote or share, which candidates with little chance of winning the election frequently post, were unpopular in terms of user engagement (Section 6).
    \item Candidates who are more likely to win use social media for broadcasting rather than for iterating with voters. 
    Also, unlike the findings of existing studies, candidates in a state of close competition have fewer interactions with other users (Section 7).
\end{itemize}

To the best of our knowledge, this is the first study to examine candidates' behavior in social media according to the odds of winning an election.
Our study can benefit from exploring how variations in candidates' situations may directly affect the way candidates engage with voters through social media.
We can gain insight into how each candidate faces social media communications in an election context.

\section{Related Work}
\subsection{Political communication on social media}
The impact of social media on political communication has been a significant topic~\cite{haq2020survey}.
Before the advent of social media, political communication based on democracy was mainly performed to be mediated by traditional mass media.
The emergence of social media has profoundly changed the form of political communication by providing new spaces for conversation and social interaction~\cite{papacharissi2004democracy}.
This change brought benefits such as revitalizing political debate and increased diversity~\cite{chadwick2008web}.
On the other hand, social media also brings negative aspects, such as political filter bubbles or echo chambers~\cite{barbera2015tweeting}.
Despite the advantages and disadvantages of the emergence of social media, there is no doubt that it is currently the most important source of political information for voters and an important forum for political actors to disseminate information~\cite{knoll2020social,pew_research}.

Much research has been conducted on how public users participate in political communication on social media~\cite{stier2020populist,blassnig2019populist}.
In particular, public users with a strong voice in the domain of politics have received attention in studies of political behavior.
Even though they are not politicians, they have many followers and influence the behavior of other users~\cite{bode2016politics,vaccari2015follow}.
On the other hand, it is said that their views do not necessarily represent the views of the groups to which they belong~\cite{barbera2015understanding}.

For political actors, the use of social media plays an important role because it can achieve various purposes; they can inform a broader belief, interact with voters, or mobilize their followers~\cite{magin2017campaigning}.
They attempt to gain support and spread their claims through political communication on social media platforms~\cite{klinger2015emergence}.
They also make an attempt to get a lot of user engagement, e.g., liking, commenting, and sharing with other users, for success in social media communication~\cite{popa2020informing}.
A large amount of user engagement depends on many factors; profile characteristics ~\cite{keller2018followers,vaccari2015follow}, the post content~\cite{xenos2017understanding}, the sentiment and style characteristics~\cite{blassnig2021popularity,heiss2019drives}, the attached images~\cite{farkas2021images}, and polarization rhetoric~\cite{ballard2022dynamics}.
Political actors are (consciously or unconsciously) concerned about what content to show and how to show it, in order to gain user engagement.

Differences in political communication in social media also emerge depending on the position and affiliation of political actors; political party~\cite{keller2018followers,blassnig2021popularity}, right-wing or left-wing~\cite{morstatter2018alt}, political leader~\cite{vaccari2015follow,jain2021twitter}, populist or not~\cite{bobba2019social,blassnig2019populist}, and famous or not~\cite{graham2013between}.
These studies have shown that position and affiliation are strongly related to the content of posts, the ease of gaining engagement, and the manner of reply.
However, it is still unclear how political communication on social media depends on the chance of winning an election.

\subsection{Election campaigns}
Elections are the prime of democracy.
Elections tend to increase the volume of posts related to politics and are the catalyst for political discussion on social media~\cite{ahmed2014my,jungherr2016twitter}.
Most political actors are naturally interested in the outcome of elections.
Even in the field of research, social media in the role of the social sensor has been considered as an alternative to polls or a possible predictor of election outcomes~\cite{tumasjan2010predicting,kulshrestha2017politically,burnap2016140}.
However, it is currently considered difficult to predict the outcome of elections from social media because of the complex relationship between post engagement and whether to win or not~\cite{jungherr2017digital}.

The political discussion among politicians and general users during the election period is very active, making it a useful subject for analysis.
While most studies have focused on presidential and congressional elections in the U.S.~\cite{paul2017compass,bovet2018validation}, some studies analyze the relationship between social media and elections in countries outside the U.S. because the relationship between social media and political communication is similar for elections in most countries; U.K.~\cite{burnap2016140}, Germany~\cite{jurgens2011small}, Belgium~\cite{boireau2014determining}, Egypt~\cite{elghazaly2016political}.
Our study focuses on the political communication of candidates in Japanese elections, same as \cite{yoshida2018information, usui2018analysis}.

\section{Data}
\subsection{2022 Japan Election}
We employ tweet data of candidates running for the 2022 Upper House of Councilors Election in Japan, which was announced on June 22 and held on July 10, 2022, to elect 125 members of the upper house of the National Diet, as the subject of our analysis.
There are some reasons why the election is a desirable case study for the aim.
First, in the 2022 elections, candidates can choose between two election ways to run; a proportional representation system or a constituency system in each prefecture.
A proportional representation system reflects the overall distribution of public support for each political party and ensures minority groups have a measure of representation proportionate to their electoral support.
A constituency system selects one or several representatives, depending on the size of the electoral district, in proportion to the number of individual votes for candidates.
In effect, two electoral ways are run in parallel during one election period, where we allow analyzing social media strategies from various perspectives.
Second, Twitter is quite popular in Japan with approximately 60 million users, and is roughly the same number of daily active users as the U.S.~\cite{nhk_news}.
It is said that 83\% of the candidates in the election also campaigned through Twitter, which is a higher percentage than in any other social media service~\cite{senkyo_dot}.
For these reasons, looking at political communications on Twitter in Japan is useful in terms of post volume and multidimensional analysis, when analyzing social media strategies during an election term.

The result of the election is that the ruling Liberal Democratic Party (LDP) increased its seats, and the largest percentage of women have been elected so far.
It is interesting to note that a new party, the Trump-inspired Party (Sanseito), won seats.
Two days before the election, the assassination of the previous prime minister Shinzo Abe caused a great flutter.

\begin{table}[t]
    \centering
    \small
    \caption{Basic stats of four groups; almost Winning (W), Even (E), almost Losing (L), and Proportional Representation (PR) group.
    It shows the number of candidates, the percentage of those winning the election, and the percentage belonging to the ruling parties. The number in parentheses is the value when including users who do not have a Twitter account.}
    \begin{tabular}{crrr} \toprule
    \multirow{2}{*}{Group} & \multicolumn{1}{c}{Number of} & \multicolumn{1}{c}{\% of winning} & \multicolumn{1}{c}{\% of the} \\ 
    & \multicolumn{1}{c}{the candidates} & \multicolumn{1}{c}{the election} & \multicolumn{1}{c}{ruling party}\\ \midrule
    W & 60 (63) & 98.33\% (98.41\%) & 65.00\% (65.07\%)\\ 
    E & 23 (23) & 47.82\% (47.82\%) & 39.13\% (39.13\%)\\ 
    L & 217 (281) & 0.92\% (0.71\%) & 1.38\% (1.06\%)\\ 
    PR & 124 (178) & 32.23\% (28.09\%) & 25.00\% (28.08\%)\\ 
    \bottomrule
    \end{tabular}
    \label{basic_info}
\end{table}

\subsection{Data collection}
We identified the Twitter accounts of the candidates in the 2022 House of Councilors Elections.
In all, 433 of the 545 candidates had Twitter accounts, and we crawled their user profiles and their posts.
We collect their data using Twitter Academic API~\cite{pfeffer2022sample}.
We treat their tweets as the subject of analysis, which are divided into two term periods based on the date of the election announcement date; the term from the election announcement to the election date called as \textit{Election} term and two months prior to the date of the election announcement called as \textit{Pre-election} term, where there is an almost similar number of tweets as during the \textit{Election} term.

\begin{figure*}[t]
  \begin{subfigure}[b]{0.24\linewidth}
    \centering
    \includegraphics[width=\textwidth]{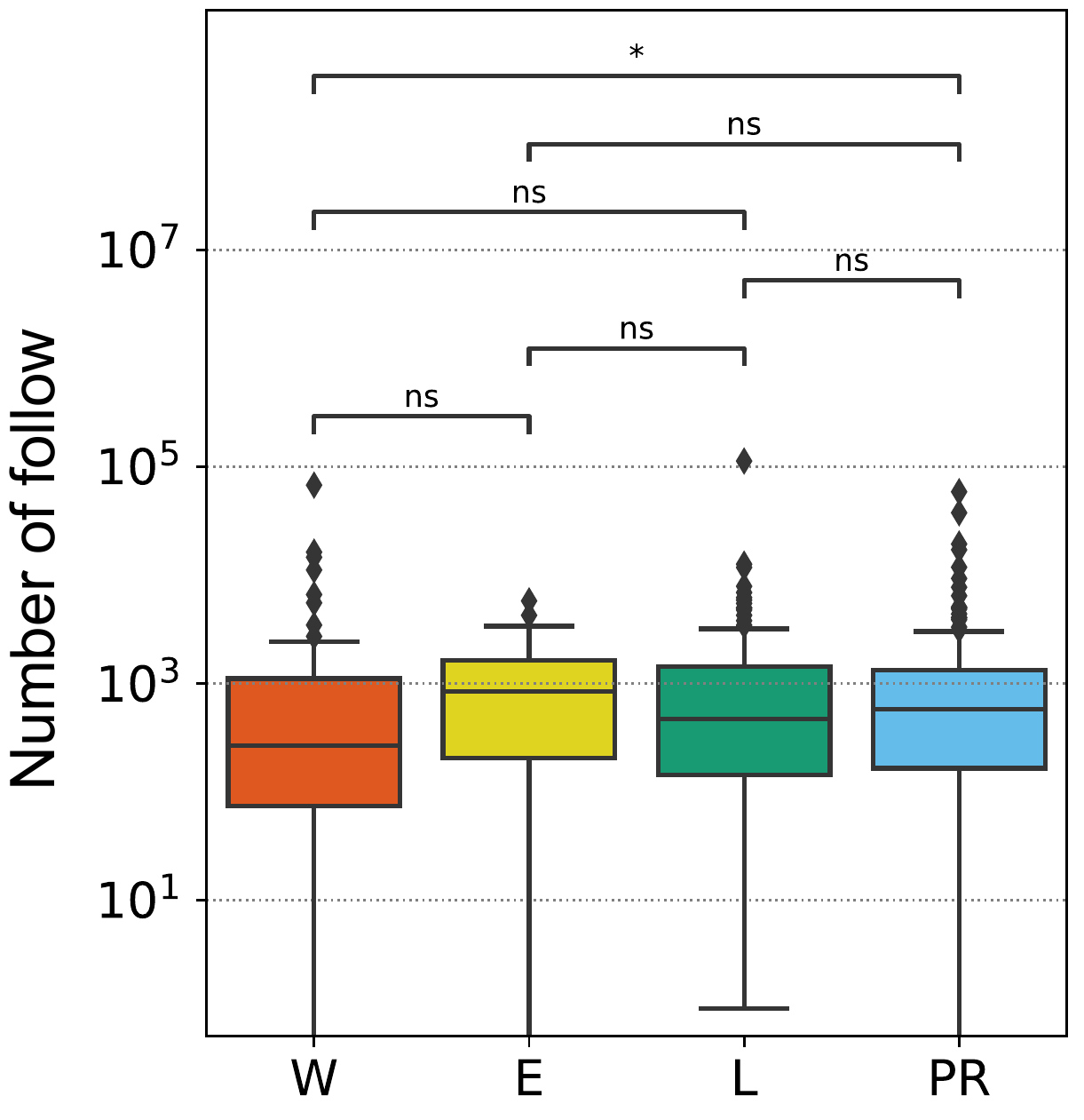}
   \subcaption{Number of following}
   \label{user_character_a}
  \end{subfigure}
  \hfill
  \begin{subfigure}[b]{0.24\linewidth}
    \centering
    \includegraphics[width=\textwidth]{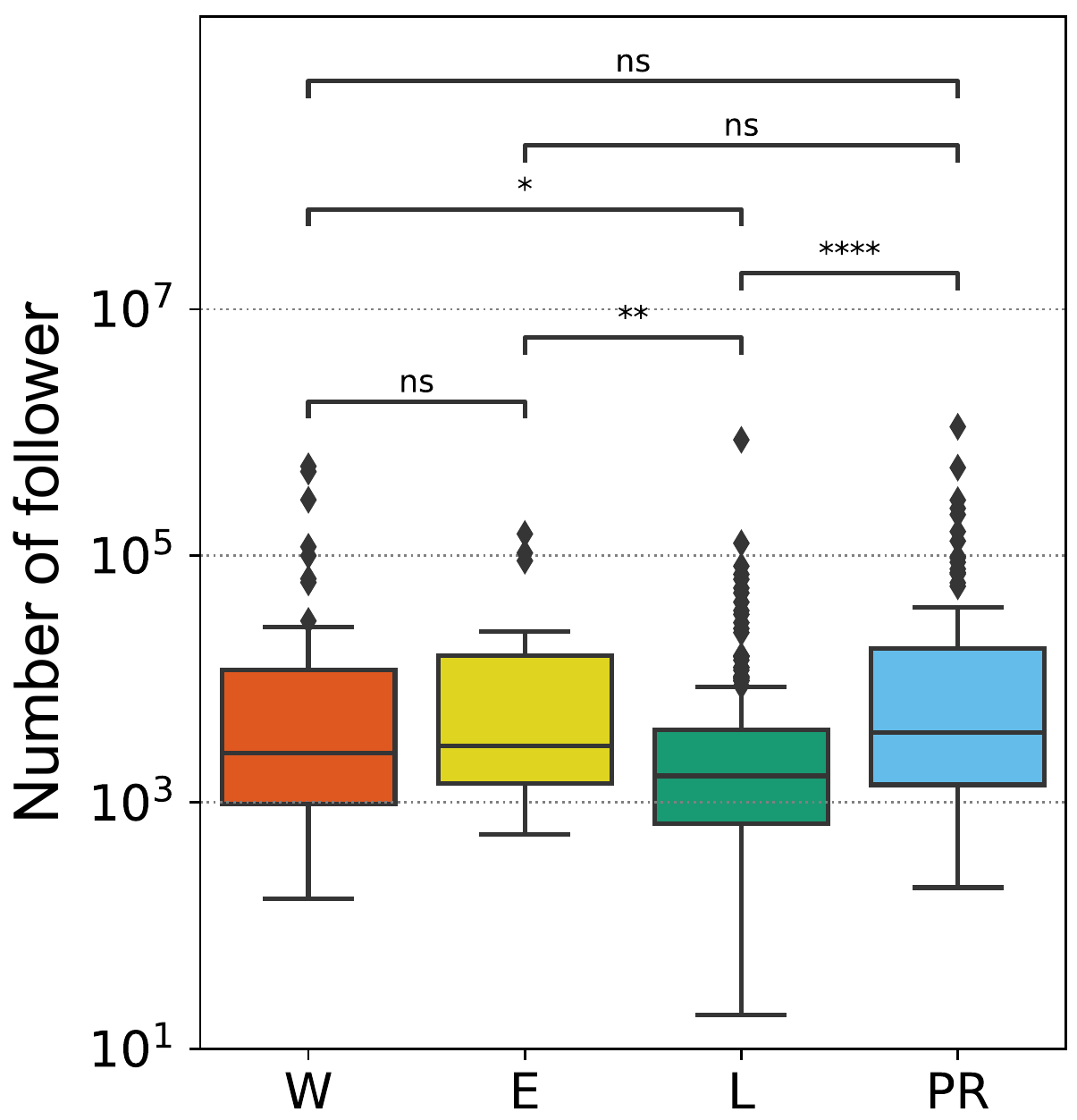}
    \subcaption{Number of followers}
    \label{user_character_b}
  \end{subfigure}
  \hfill
  \begin{subfigure}[b]{0.24\linewidth}
    \centering
    \includegraphics[width=\textwidth]{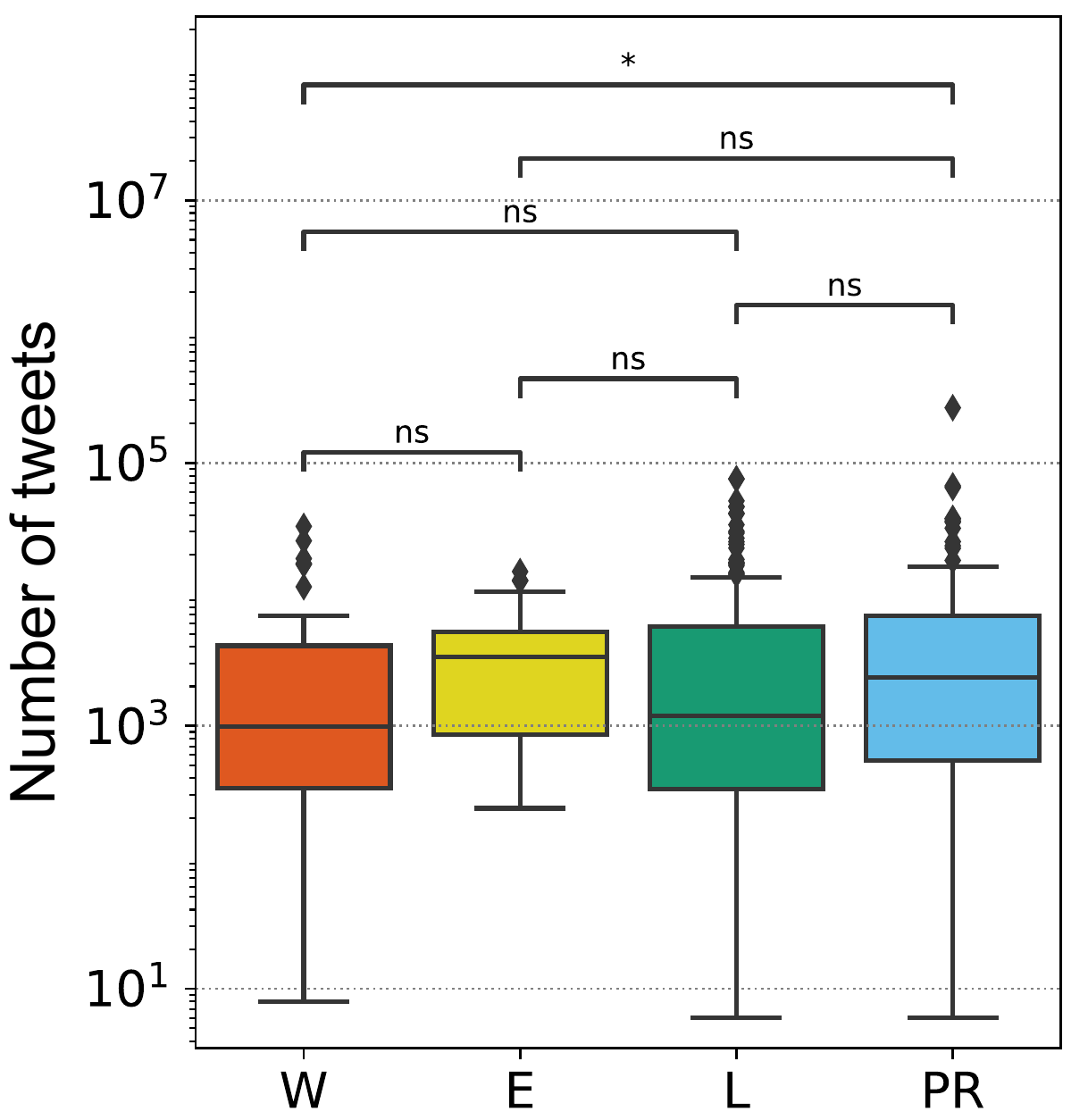}
    \subcaption{Number of tweets}
    \label{user_character_c}
  \end{subfigure}
  \hfill
  \begin{subfigure}[b]{0.24\linewidth}
    \centering
    \includegraphics[width=\textwidth]{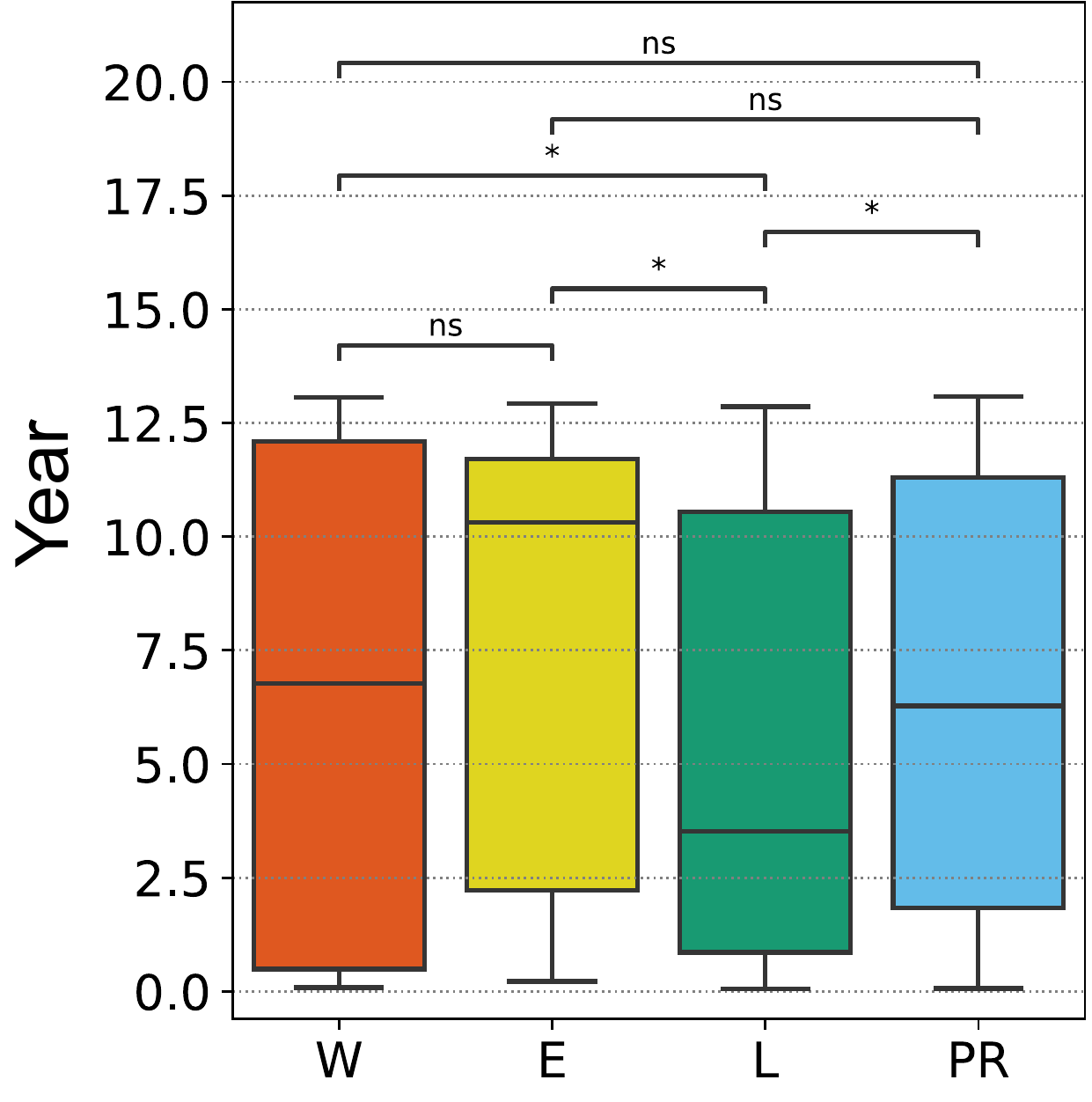}
    \subcaption{Account age}
    \label{user_character_d}
  \end{subfigure}
  \caption{
  Comparison of the four types of user information for each group represented by box plot; (a) Number of following, (b) Number of followers, (c) Number of tweets, and (d) Account age.
  We test whether two groups are likely to derive from the same group using Mann Whitney U (MWU) tests.
  We express the degree of statistical significance in stars; ns: p-value $\leq 1.00$, *:  p-value $\leq 0.05$, **:  p-value $\leq 0.01$, ***:  p-value $\leq 0.001$, and ****:  p-value $\leq 0.0001$}
  \label{user_character}
\end{figure*}

Our study focuses on how political communication on social media differs depending on their chances of winning the election.
We divided the candidates' accounts into four groups; almost Winning (W), Even (E), almost Losing (L), and Proportional Representation (PR) group.
In other words, constituency candidates assign to three groups depending on their chances of winning and proportional-representation candidates belong to the fourth group because whether they win the election depends largely on the popularity of the party to which they belong.
The judgment criterion for the assignment of the groups of each candidate was based on the survey of the electoral situation published by the Asahi Shimbun, which is the third largest newspaper in the world, three days after the announcement of the election~\cite{asahi_shin}.
Their surveys are reported whether each candidate takes the lead in the election or not, based on their coordination of the situation and interviews.
We assign the candidates judged to be superior or slightly superior to the ``almost Winning (W)'' group, those judged to be in a state of close competition to the ``Even (E)'' group, and those judged to be inferior or no mention to the ``almost Losing (L)'' group.

The basic stats are shown in Table~\ref{basic_info}.
The percentage of Groups W and E with Twitter accounts is near 100\%, while those of Groups L and PR are 77.2\% and 69.6\%, respectively.
Groups that are likely to win or are unsure of their chances of winning appear to be more active in engaging in social media strategy.
Also, since the percentage of candidates for each group that won the election decreases in order from group W to group L, the survey as the basis for the group assignment appears to be reasonable.

What is the relationship between group assignments and the ruling and opposition parties?
The ratio of ruling parties shows an extreme bias in the ``Almost Losing (L)'' group.
This is due to the fact that the ruling Liberal Democratic Party (LDP) is very selective in its candidates, and some opposition parties are fielding large numbers of candidates.
The ``Almost Losing (L)'' group is mostly composed of members in the opposition parties, but the other groups are unbiased and mixed with both the ruling and opposition parties.
Note that our study is based on the perspective of the chance of winning an election, and offers a new perspective on political communication on social media, that differs from the dichotomy perspective between the ruling party and the opposition.

\section{RQ1: What are the characteristics of the frequency of tweets and user information?}
This section characterizes the frequency of tweets and user information by the four groups we have defined to understand the differences among their strategies in social media.

\subsection{User characteristics}
We examine the differences among the groups for four types of user characteristics; the number of following, the number of followers, the number of tweets, and account age. 
The results are shown in Figure~\ref{user_character}.
The comparison is expressed by box plot, and we use Mann Whitney U tests~\cite{mann1947test} for the statistical significance test.

The number of following of candidates does not differ significantly between any of the groups, except the pair of W and PR, as shown in Figure~\ref{user_character} (a).
Although group E appears to tend to have fewer followers than the other groups, it is apparent that the majority of users only follow 100 -- 1,000 users.
The number of followers shows that group L was significantly different from the other groups, as shown in Figure~\ref{user_character} (b).
The lower bar of the box plot in group L is spreading downwards, suggesting that some of the users in group L have not fully gained popularity on social media.
For example, the median number of followers in group W is 31,835, while that in group L is 9,849.
On the other hand, several candidates in group L have more than 100,000 followers.
This implies that while more followers (the popularity on social media) does not necessarily increase the chance of winning an election, above a certain amount of followers is necessary to have a certain chance of winning, i.e., to join group W or E.

\begin{figure}[t]
    \centering 
    \includegraphics[width = \linewidth]{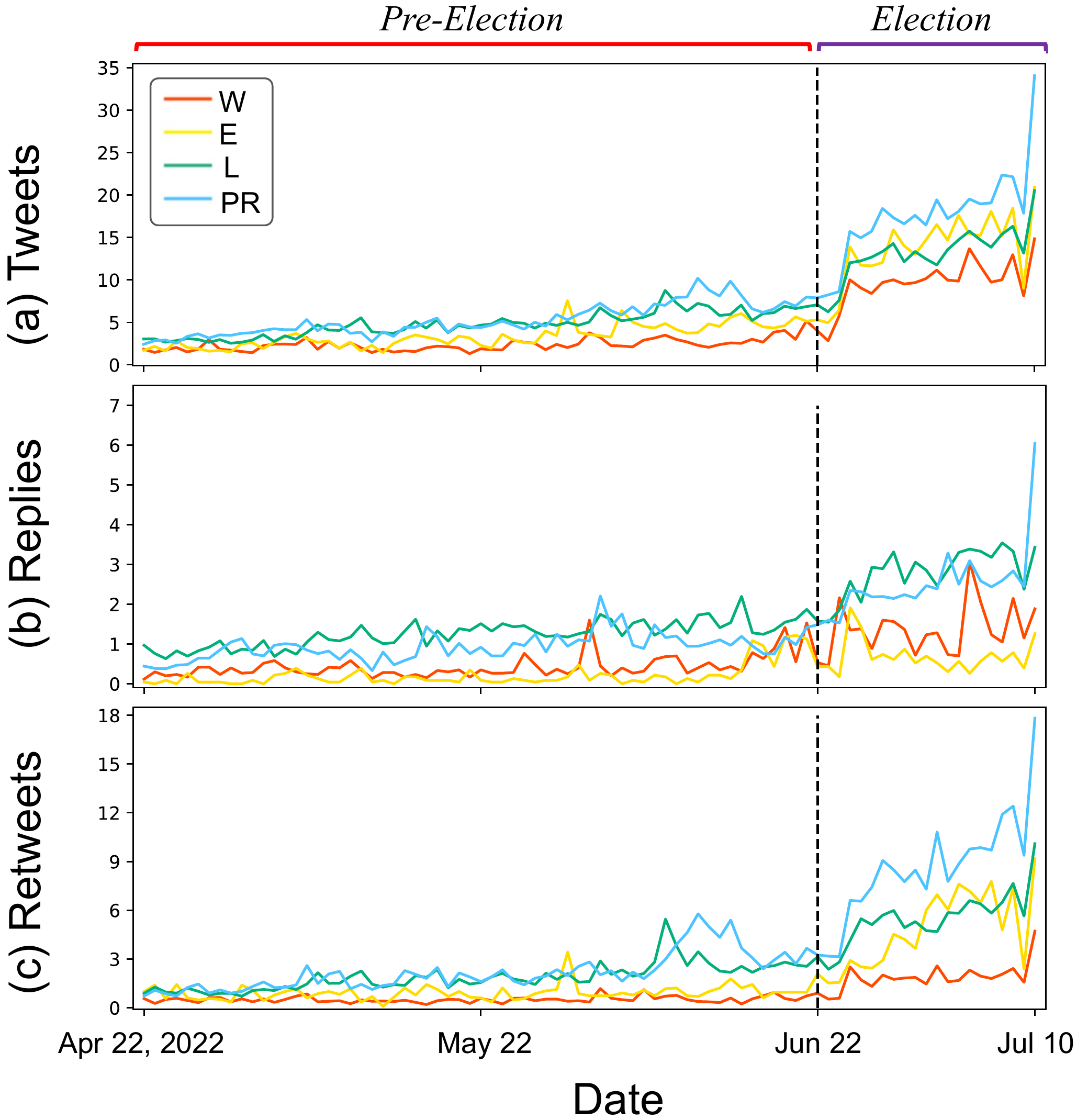}
    \caption{The average number of (a) all tweets, (b) replies, and (c) retweets from Apr 22 to Jul 10, 2022.
    Each colored line represents each group.
    We set two months prior to the date of the election announcement as \textit{Pre-election} term and the term from the election announcement to the election date as \textit{Election} term.
    }
    \label{time_series_tweets}
\end{figure}

For the number of tweets, the result in Figure~\ref{user_character} (c) shows that W and PR are significantly different, the same as the number of following.
It indicates that the facing of social media differs among the leading candidates depending on the election way.
Both the number of followers and the number of tweets are higher for the PR group to which candidates running in the proportional representation system belong, showing that they differ in their engagement with other public users.
Taking into account age, group L is significantly different from the other groups, as shown in Figure~\ref{user_character} (d).
Both the number of followers and account age in group L are smaller than in other groups due to the fact that there is some first-time candidate in the group.
Also, the number of tweets in group L is almost the same as those in the other groups despite the young age of the account, indicating that they are actively working on the social media strategy.


\begin{figure}[t]
  \begin{subfigure}[b]{0.32\linewidth}
    \centering
    \includegraphics[width=\textwidth]{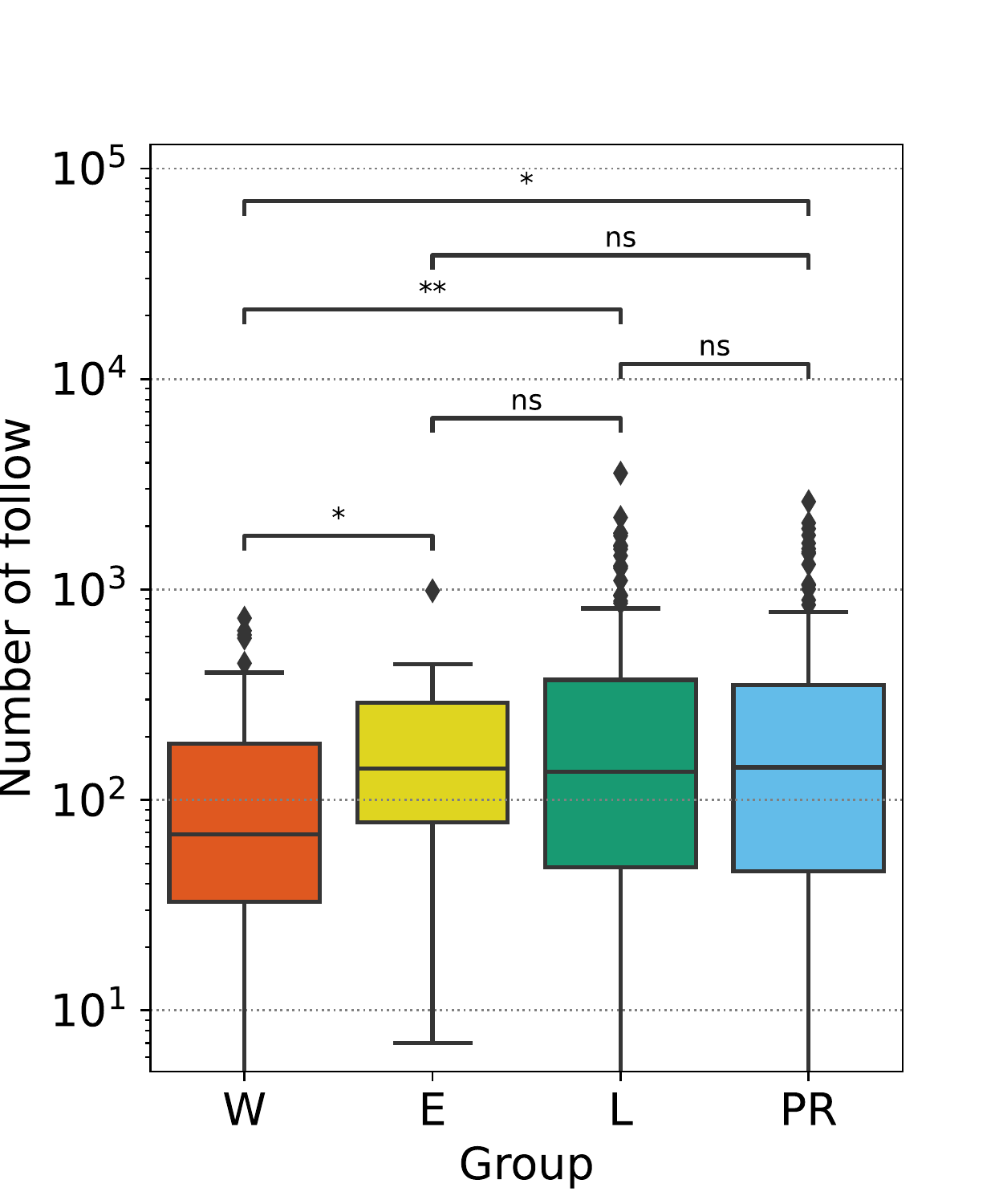}
    \subcaption{\scriptsize{Number of tweets}}
  \end{subfigure}
  \hfill
  \begin{subfigure}[b]{0.32\linewidth}
    \centering
    \includegraphics[width=\textwidth]{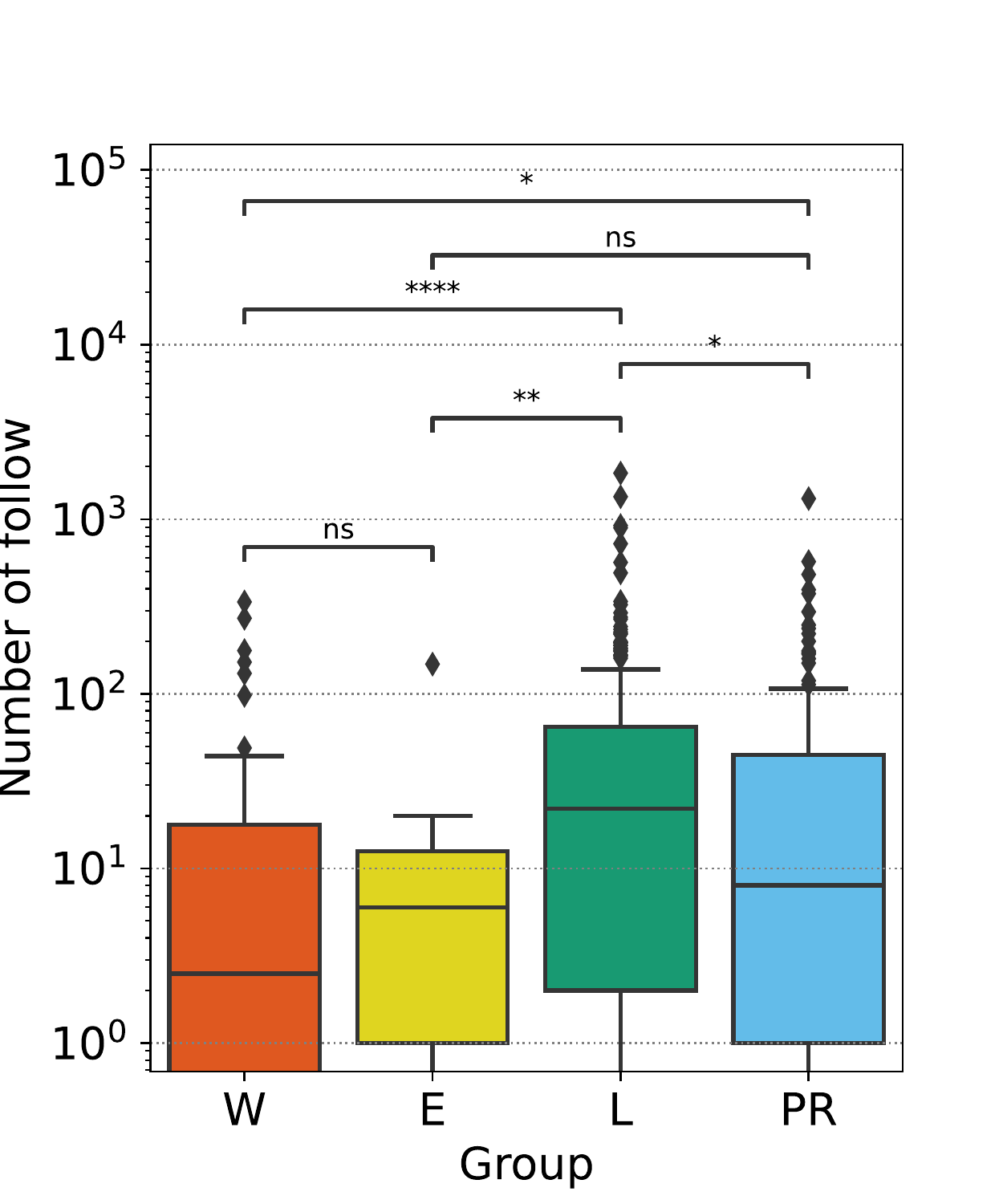}
    \subcaption{\scriptsize{Number of replies}}
  \end{subfigure}
  \hfill
  \begin{subfigure}[b]{0.32\linewidth}
    \centering
    \includegraphics[width=\textwidth]{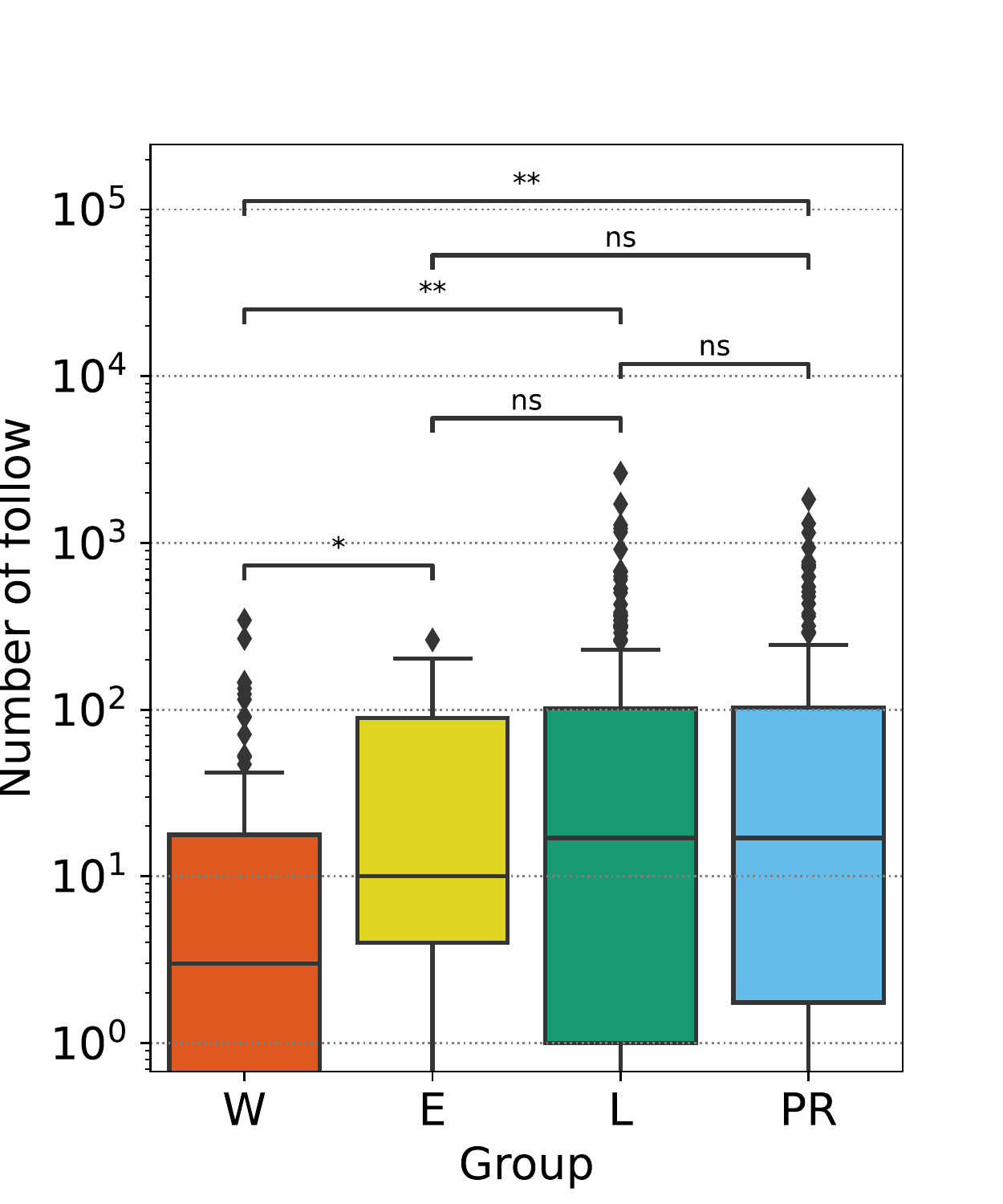}
    \subcaption{\scriptsize{Number of retweets}}
  \end{subfigure}
  \caption{Number of tweets for each group during \textit{Pre-election} term.}
  \label{pre_election_tweet}
\end{figure}

\begin{figure}[t]
  \begin{subfigure}[b]{0.32\linewidth}
    \centering
    \includegraphics[width=\textwidth]{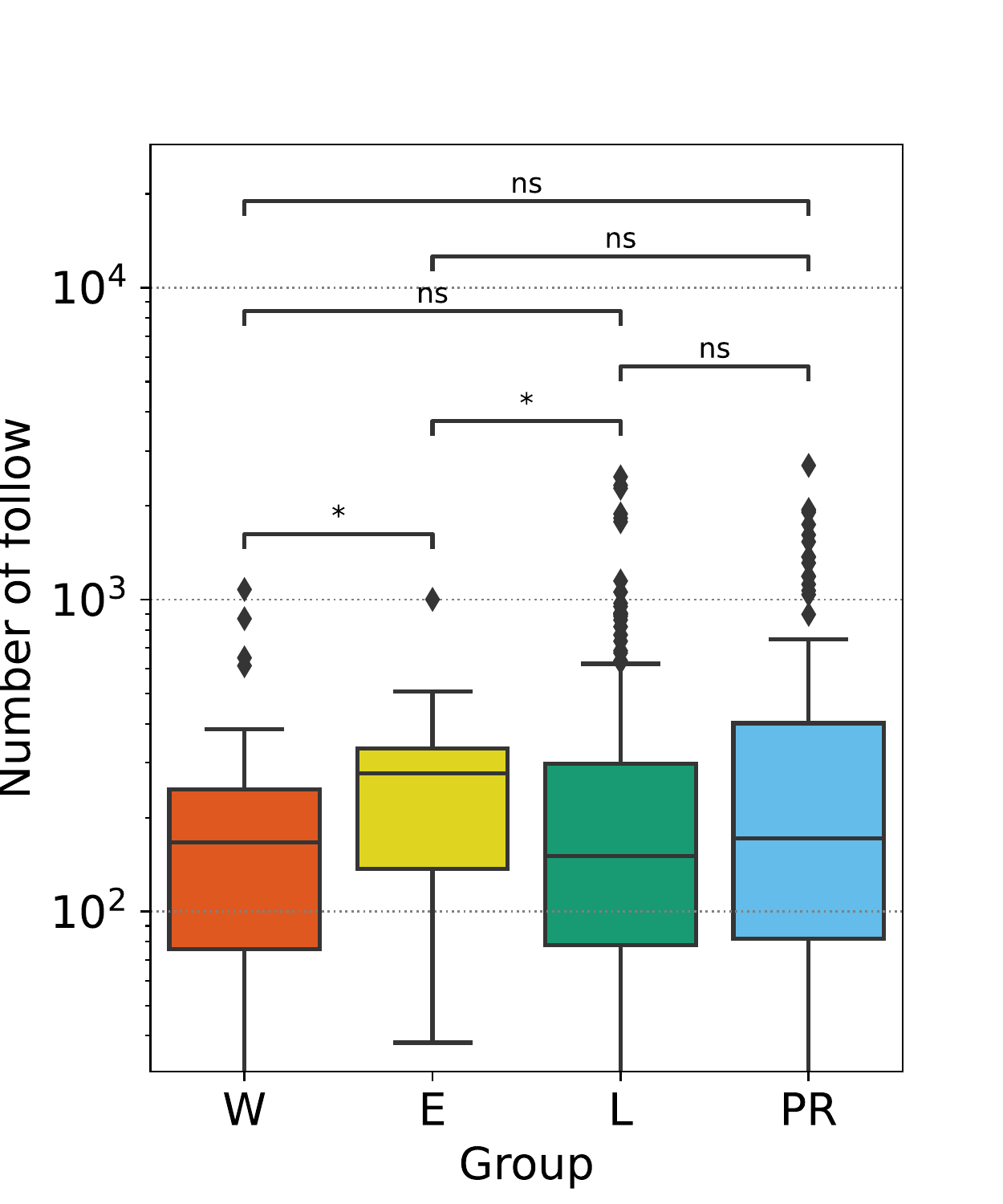}
    \subcaption{\scriptsize{Number of tweets}}
  \end{subfigure}
  \hfill
  \begin{subfigure}[b]{0.32\linewidth}
    \centering
    \includegraphics[width=\textwidth]{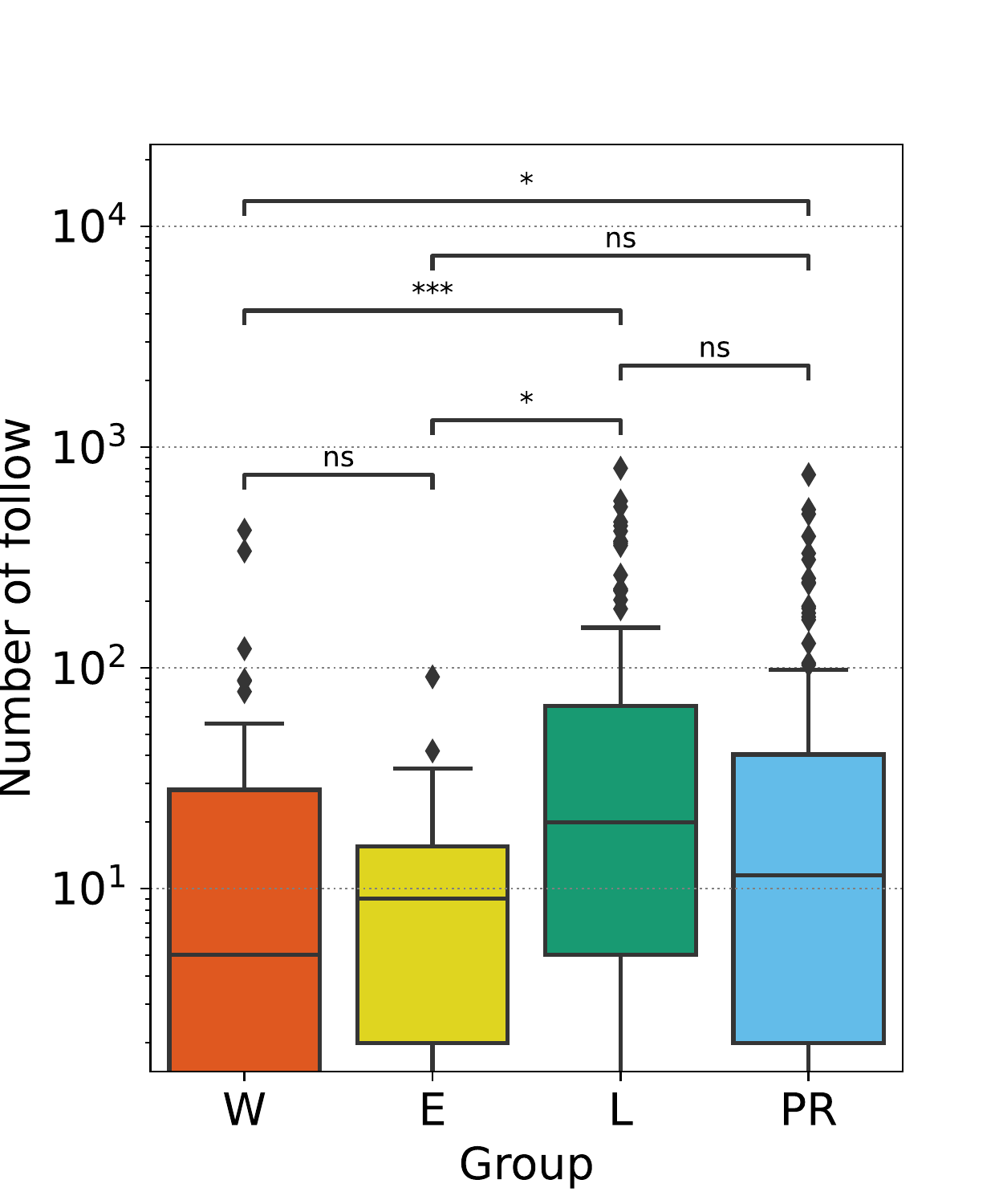}
    \subcaption{\scriptsize{Number of replies}}
  \end{subfigure}
  \hfill
  \begin{subfigure}[b]{0.32\linewidth}
    \centering
    \includegraphics[width=\textwidth]{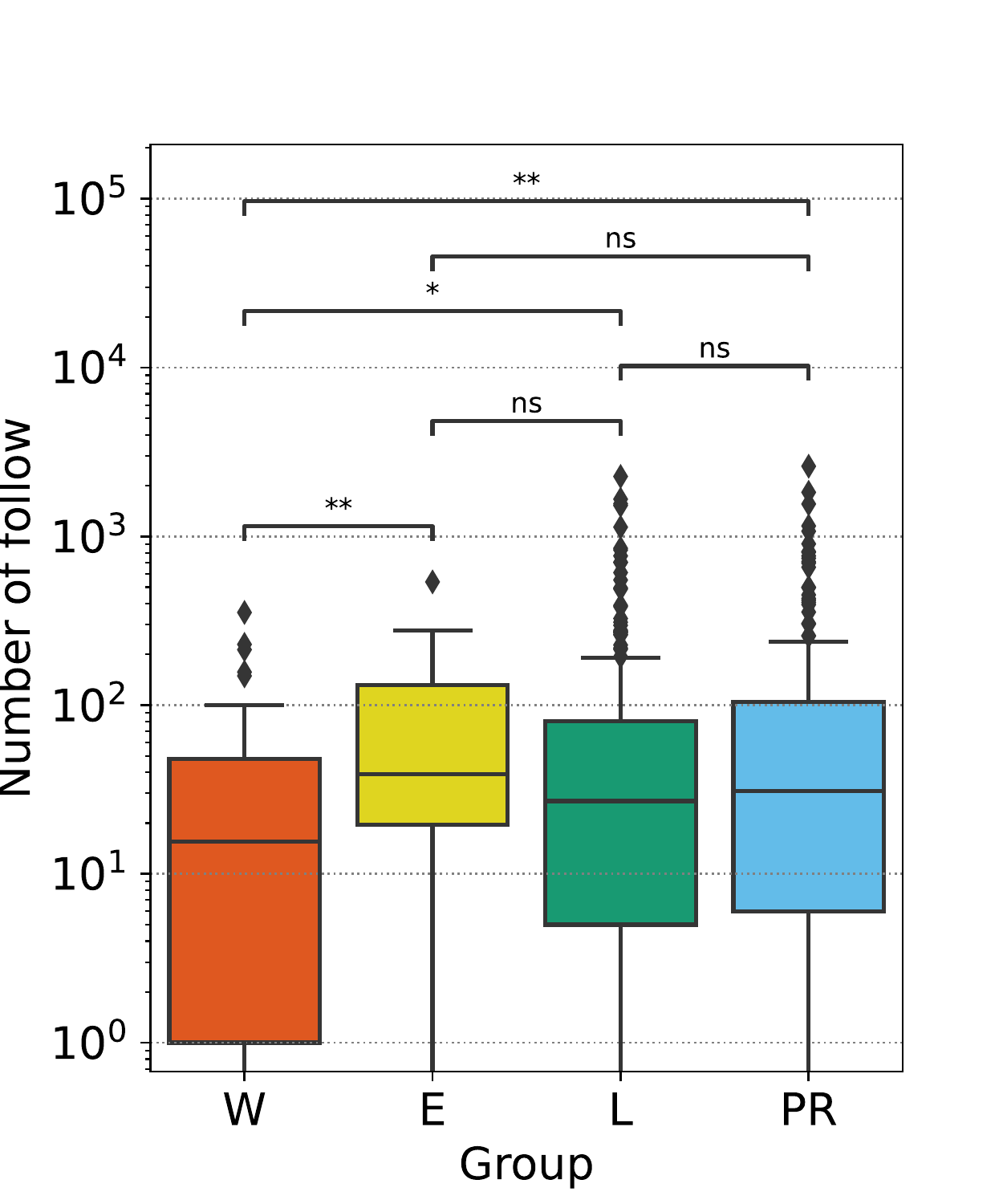}
    \subcaption{\scriptsize{Number of retweets}}
  \end{subfigure}
  \caption{Number of tweets for each group during \textit{Election} term.}
  \label{election_tweet}
\end{figure}

\subsection{Time series in a number of tweets}
We investigate whether there is a difference in the number of tweets by each group during \textit{Pre-Election} and \textit{Election} terms.
The time series of the average number of tweets are shown in Figure~\ref{time_series_tweets}, and the comparisons of the number of tweets in each term are shown in Figures~\ref{pre_election_tweet} and \ref{election_tweet}.

The number of tweets during \textit{Election} term tends to increase more than those during \textit{Pre-Election} term because of the activation of electoral campaigns on social media.
During \textit{Pre-Election} term, the number of tweets, replies, and retweets is significantly fewer for group W than for other groups.
Candidates who have already gained popularity use social media in smaller amounts.
On the other hand, group L makes use of replies more than any other group during the \textit{Pre-Election} term, indicating that it is aware of the dialogue with other users even before the election.

During \textit{Election} term, group W significantly posts fewer tweets, replies, and retweets, which is similar characteristics as during \textit{Pre-Election} term.
Group E tends to have larger tweets but fewer replies than other groups.
It suggests that the competitive state of their rivals has led to an increase in their own election campaign tweets.
Group L focuses on interacting with other users through a reply function, the same as during \textit{Pre-Election} term.
Interestingly, group PR retweets during \textit{Election} term more than other groups.
We consider that the candidates in group PR frequently retweet their party's propaganda because the rise in popularity of their party directly leads to their electoral triumph, due to the electoral system in which they run.

\section{RQ2: What kind of content does each group post during the election period?}
In this section, to better understand the topics to which each group tends to refer, we use a topic modeling approach to group tweets into meaningful topics.
The identification of the content that each group talks about makes clear what election issues they want the public to pay attention to and what they want to claim.


\begin{figure*}[t]
  \begin{subfigure}[b]{0.45\linewidth}
    \centering
    \includegraphics[width=\textwidth]{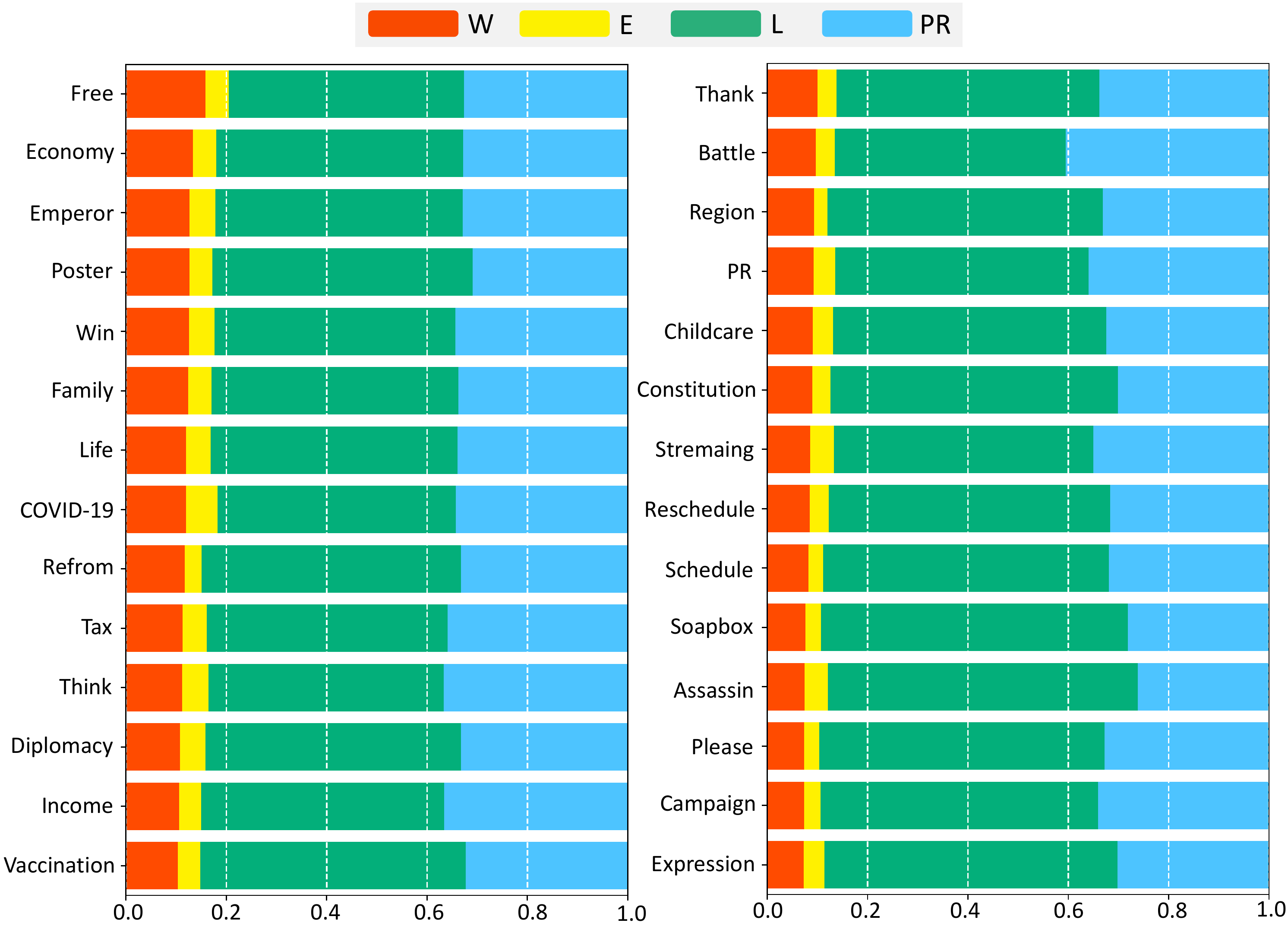}
    \subcaption{\textit{Pre-Election} term}
  \end{subfigure}
  \hfill
  \begin{subfigure}[b]{0.45\linewidth}
    \centering
    \includegraphics[width=\textwidth]{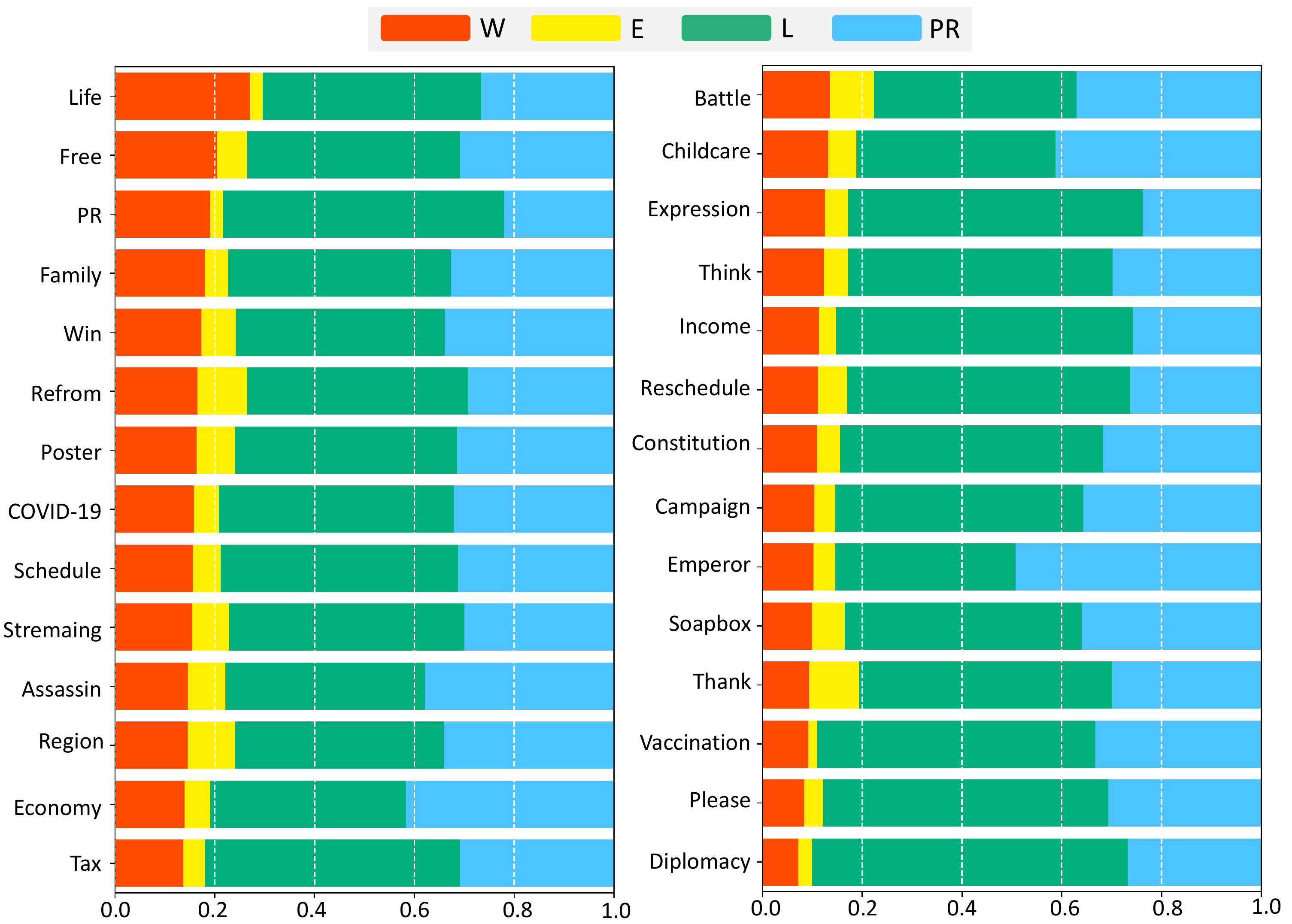}
    \subcaption{\textit{Election} term}
  \end{subfigure}
    \caption{Fraction of groups in each topic.}
    \label{fraction}
\end{figure*}

\subsection{Topic model}
We use a topic model to group all 211,495 tweets except retweets posted by candidates belonging to each group from Apr 22 to Jul 10, 2022 (in \textit{Election} and \textit{Pre-Election} terms) into clusters and to describe their properties.
We chose the Biterm Topic Model (BTM)~\cite{yan2013biterm} as our clustering method.
This model is a derivative of Latent Dirichlet Allocation (LDA)~\cite{blei2003latent} and is known to be able to extract topics with high accuracy for short sentences.
Specifically, this method assigns topics based on word sets with high co-occurrence rates among word pairs in a single tweet.

The input to BTM is word pairs for each tweet.
Before creating these word pairs, as a text preprocessing step, we remove particles, auxiliaries, and stop words, and replace some words with specific words.
Each candidate frequently uses one's name and one's party affiliation in tweets for his/her election campaign.
To mitigate the impact of individual and party names in the topic model, we replace each individual name with \textit{PERSONAL\_NAME} and the party name with \textit{PARTY\_NAME}.

As the number of clusters, we searched for the appropriate number of topics by coherence score in increments of 5 in the range of 10 to 100, and chose 35 as the initial number of topics.
We clustered the preprocessed tweets into 35 topics using BTM.
Then, we merged pairs of topics that were similar among the estimated set of topics.
Concretely, the top 50 words in each topic were extracted according to $\phi$, which represents the distribution of words in the topic, and a pair of topics with more than 20\% overlap between these words was merged as one topic.
Finally, we set 28 topics, which one of the co-authors reviewed, inspecting the words as well as the context in which they appeared, and assigned a label to each topic; Free, Economy, Emperor, Poster, Win, Family, Life, COVID-19, Reform, Tax, Think, Diplomacy, Income, Vaccination, Thank, Battle, Region, Proportional Representation (PR), Childcare, Constitution, Streaming, Reschedule, Schedule, Soapbox, Assassin, Please, Campaign, and Expression.
The percentage of each group for each topic is shown in Figure~\ref{fraction}.

\begin{table}[t]
  \caption{Topics with a large and small number of tweets in each group}
  \label{topic_number}
  \centering
  \footnotesize
  \scalebox{0.8}{
  \begin{tabular}{ccllll}  \toprule
  & Ranking & W & E & L & PR\\ \midrule 
  & 1 & Schedule & Schedule & Schedule & Schedule\\
  & 2 & Diplomacy & Diplomacy & Diplomacy & Diplomacy\\
  & 3 & Please & Please & Campaign & Campaign\\
 \textit{Pre-} & \vdots & & & & \\
 \textit{Election} & 26 & Constitution & Expression & Income & Income\\
   & 27 & Battle & Constitution & Reschedule & Reschedule\\
   & 28 & Reschedule & Win & Battle & Win\\ \midrule
  & 1 & Please & Region & Please & Please\\
  & 2 & Schedule & Diplomacy & Campaign & Diplomacy\\
  & 3 & Streaming & Streaming & Diplomacy & Schedule\\
  \textit{Election} & \vdots & & & & \\
  & 26 & Childcare & Constitution & Childcare & Childcare\\
   & 27 & Income & Childcare & Economy & Income\\
   & 28 & Constitution & Income & Income & Constitution\\ \bottomrule
  \end{tabular}
  }
\end{table}

\subsection{Results}
First, we examine the number of tweets belonging to each topic to identify popular and unpopular topics.
In all tweets, the topic with the most tweets is Please (11.88\%), which includes tweets asking users to do something please (e.g., vote), and the topic with the fewest tweets is Constitution (0.87\%) related to constitutional amendments.
The results of topics with high and low numbers for tweets in each group are shown in Table~\ref{topic_number}.
The topics with a large number of tweets are similar in all groups.
During \textit{Pre-Election} term, it shows a high number of tweets on four topics; Schedule, Diplomacy, Please, and Campaign.
Before the election announcement, there were many tweets on election-related topics such as Schedule, which reports the upcoming schedule, and Campaign, which reports the schedule of campaign speeches, suggesting that preparations for the election were being made early on.
In addition, the tense situation in Russia and Ukraine has led to many tweets about Diplomacy.
During \textit{Election} term, in groups W and E, tweets on Streaming with respect to television and internet broadcast increase.
The candidates with a high chance of winning increase to have opportunities to appear on TV, about which they tweet.
Topics with a small number of tweets during \textit{Pre-Election} naturally include subjects that are likely to be posted after the start of the election, such as Reschedule, Win, and Battle.
During \textit{Election} term, the number of tweets on topics that are usually central to political discussions, such as Childcare, Income, Constitution, and Economy, is few.
It is apparent that during the election period, they do not discuss political issues but focus on the promotion of themselves and their political party.

Observations of popular and unpopular topics showed no big differences between the groups.
Therefore, we introduce a new index, which is a form similar to Pearson's Chi-square statistics ($\chi^2$)~\cite{greenwood1996guide}, to discover topics that are distinctive in each group.
This index quantifies the degree to which each group deviates upward from the expected probability on each topic, which we call Dev\_Score.
It is represented by the following equation; 

\begin{align}
\alpha^{i}_{j} &= \tfrac{Tweets^{\text{group}\ i}_{\text{topic}\ j}}{\sum_{k = 1}^{\text{N\ of\ topics}}Tweets^{\text{group}\ i}_{\text{topic}\ k}} \nonumber \\
\mu^{\setminus i}_{j} &= \tfrac{1}{\text{N\ of\ groups} - 1}\sum_{k = 1; k \neq i}^{\text{N\ of\ groups}}\alpha^{k}_{j} \nonumber \\
\text{Dev\_Score}^{i}_{j} &= \tfrac{\alpha^{i}_{j} - \mu^{\setminus i}_{j}}{\mu^{\setminus i}_{j}} \nonumber
\end{align}
, where $\alpha^{i}_{j}$ represents the fraction of topic $j$ in group $i$, $\mu^{\setminus i}_{j}$ represents the average of each $\alpha^{*}_{j}$ except group $i$.
The deviation score $\text{Dev\_Score}^{i}_{j}$ indicates the degree of specificity of topic $j$ in group $i$, compared to other groups.

Table~\ref{topic_kai} shows the top three topics of Dev\_Score for each group.
Group W tends to focus on Free such as freedom of expression more than other groups.
Group E shows that the intensity of the elections triggered tweets related to the area in which they are running for office (Region).
They tweeted extensively on Streaming about their own appearances on TV and the Internet to increase their visibility and attention.
In addition, in the \textit{Election} term, they posted more on Assassin, the topic of tweets about former Prime Minister Abe's assassination, than any other group.
This indicates a tendency to mention sensational incidents.
Group L actively posted Please tweets, urging people to vote and support the campaign in both terms, suggesting that the campaign is in a tough state.
In group PR, during \textit{Election} term, they often tweet about the election, such as Please and Battle.
While, during \textit{Pre-Election} term, they are likely to mention political policies such as Reform and Childcare or express their own ideas (Think), suggesting that they are less aware of the election before the election than other groups because of the largely dependent on the popularity of their political party about winning the election.

\begin{table}[t]
  \caption{Top 3 topics for the Dev\_Score for each group}
  \label{topic_kai}
  \centering
  \footnotesize
  \scalebox{0.8}{
  \begin{tabular}{ccllll}  \toprule
  & Ranking & W & E & L & PR\\ \midrule 
  \multirow{2}{*}{\textit{Pre-}} & 1 & Free & Diplomacy & Please & Reform\\
  \multirow{2}{*}{\textit{Election}}& 2 & Poster & Streaming & Campaign & Think\\
  & 3 & Economy & Poster & PR & Childcare\\ \midrule
  & 1 & Life & Region & Please & Please\\
  \textit{Election} & 2 & Free & Assassin & Campaign & Battle\\
  & 3 & PR & Streaming & Expression & Emperor\\ \bottomrule

  \end{tabular}
  }
\end{table}

\section{RQ3: What type of content affects user engagement?}

\begin{figure*}[t]
  \begin{subfigure}[b]{0.24\linewidth}
    \centering
    \includegraphics[width=\textwidth]{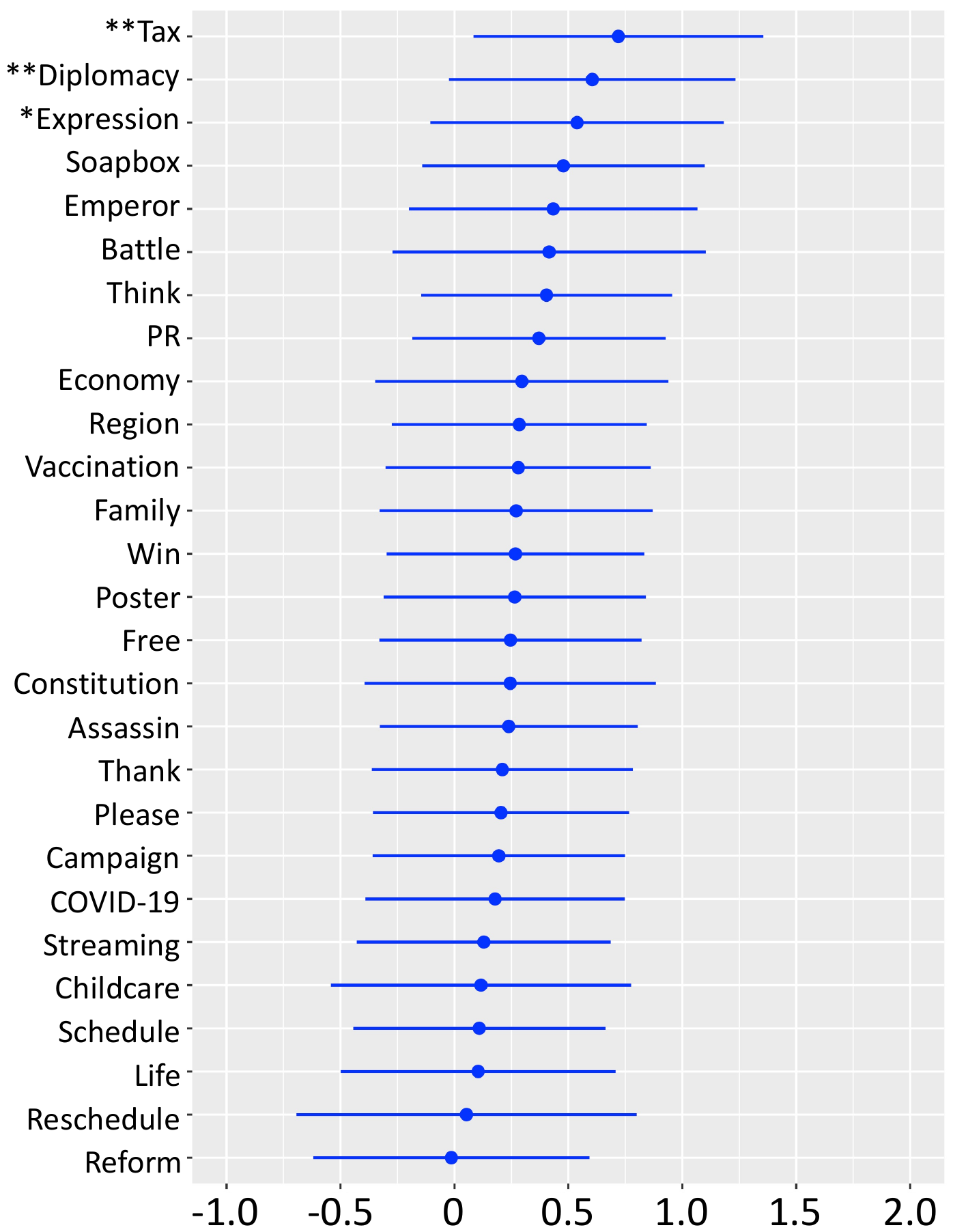}
    \subcaption{Group W in \textit{Pre-Election}}
  \end{subfigure}
  \hfill
  \begin{subfigure}[b]{0.24\linewidth}
    \centering
    \includegraphics[width=\textwidth]{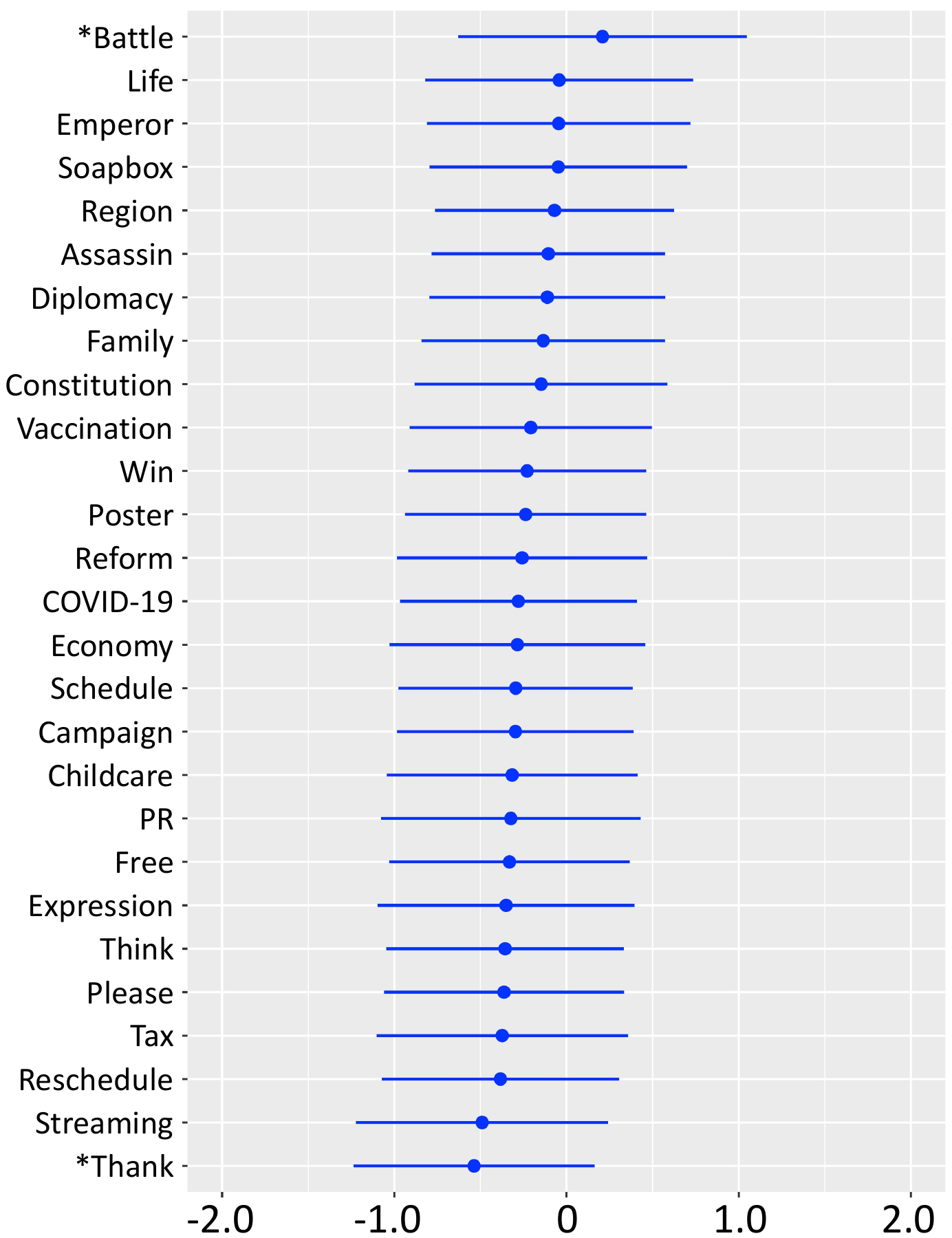}
    \subcaption{Group E in \textit{Pre-Election}}
  \end{subfigure}
  \hfill
  \begin{subfigure}[b]{0.24\linewidth}
    \centering
    \includegraphics[width=\textwidth]{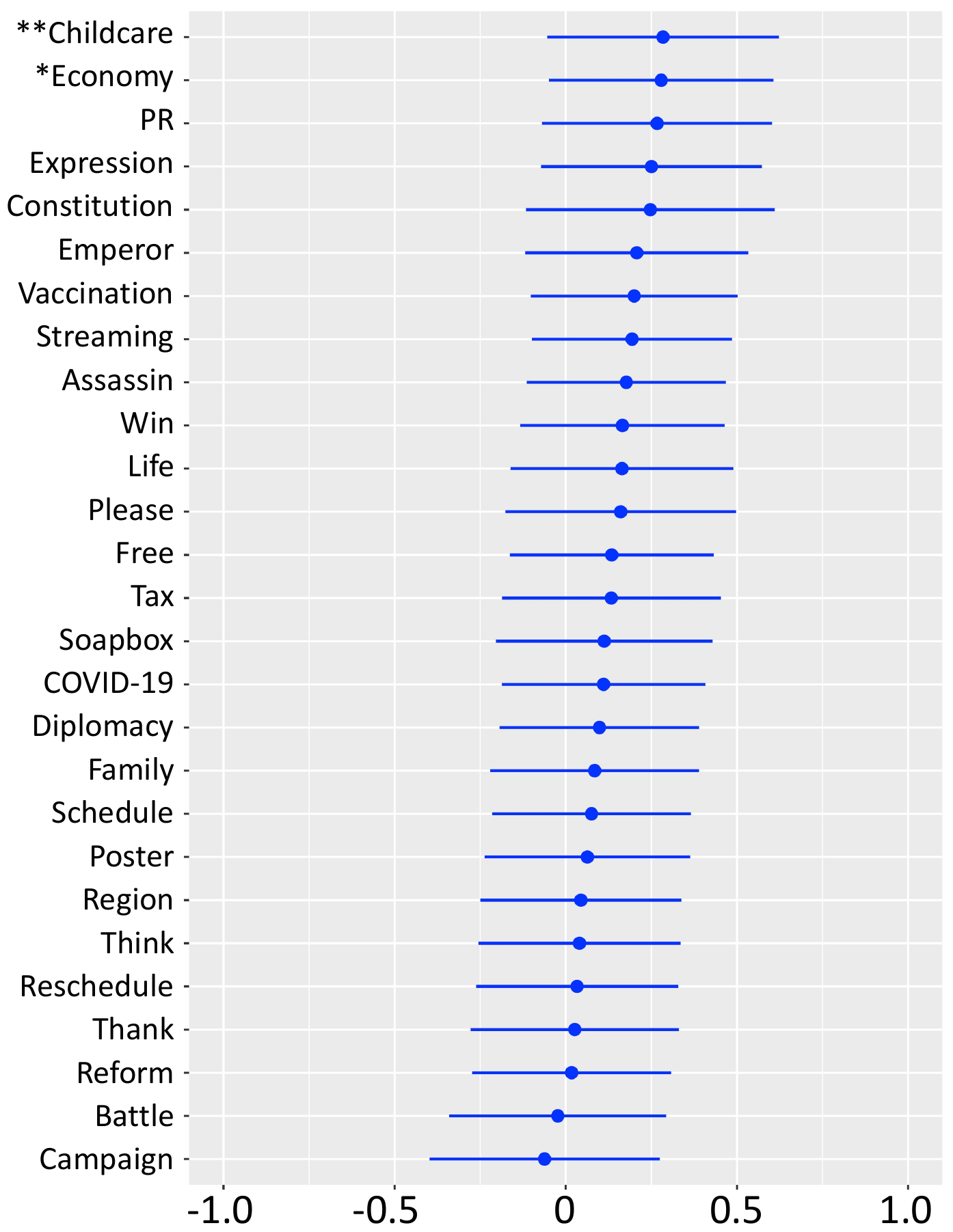}
    \subcaption{Group L in \textit{Pre-Election}}
  \end{subfigure}
  \hfill
  \begin{subfigure}[b]{0.24\linewidth}
    \centering
    \includegraphics[width=\textwidth]{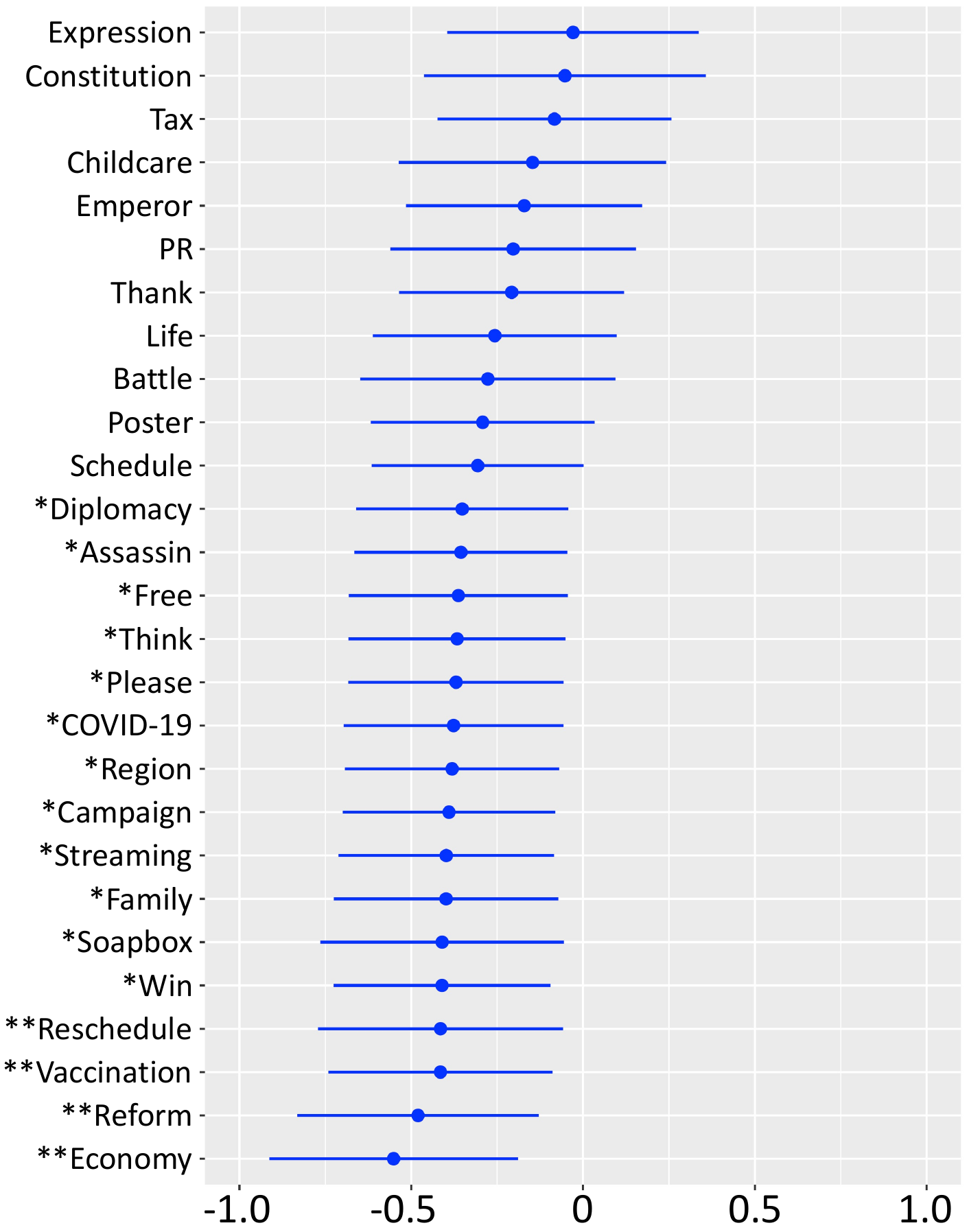}
    \subcaption{Group PR in \textit{Pre-Election}}
  \end{subfigure}
  \\
  \begin{subfigure}[b]{0.24\linewidth}
    \centering
    \includegraphics[width=\textwidth]{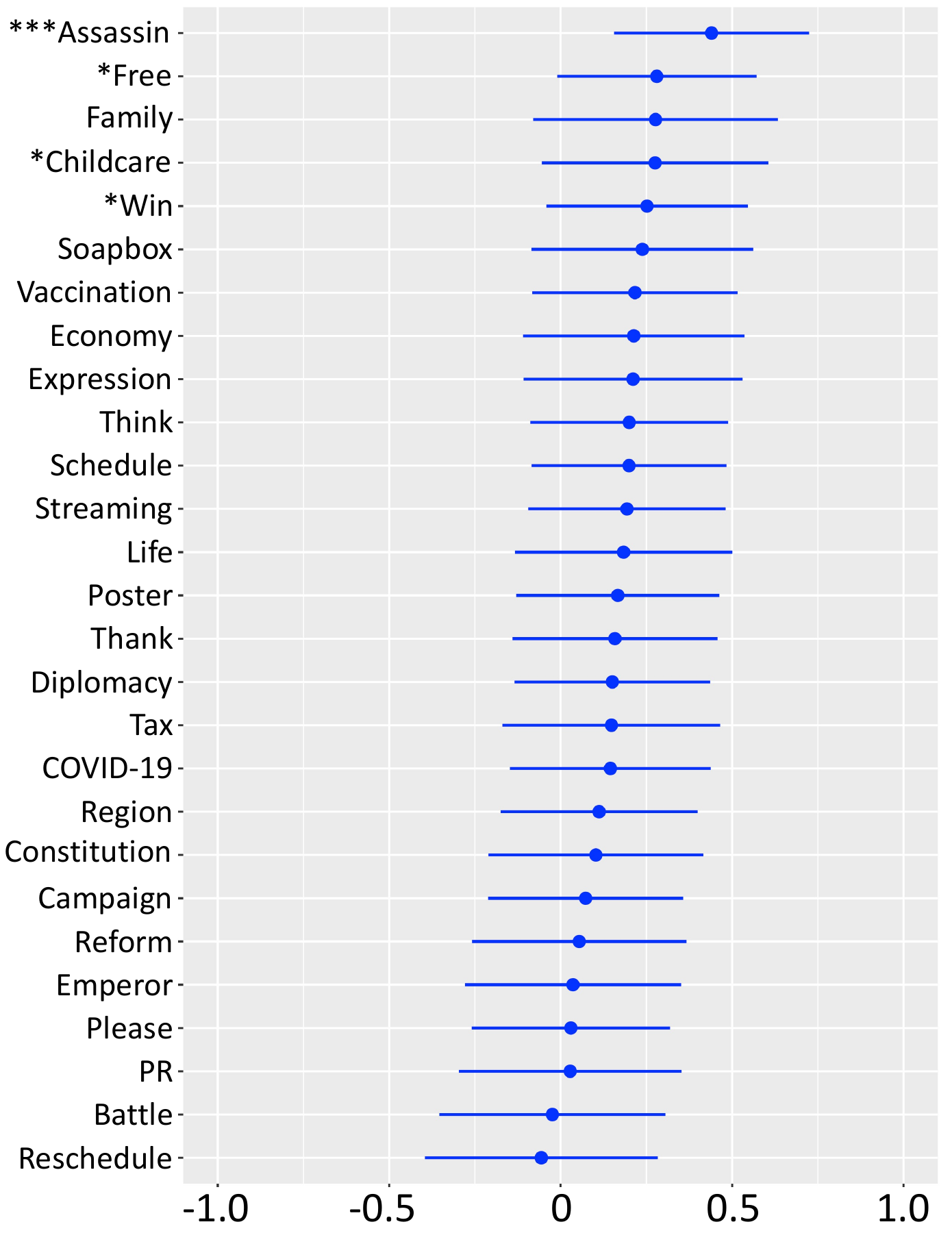}
    \subcaption{Group W in \textit{Election}}
  \end{subfigure}
  \hfill
  \begin{subfigure}[b]{0.24\linewidth}
    \centering
    \includegraphics[width=\textwidth]{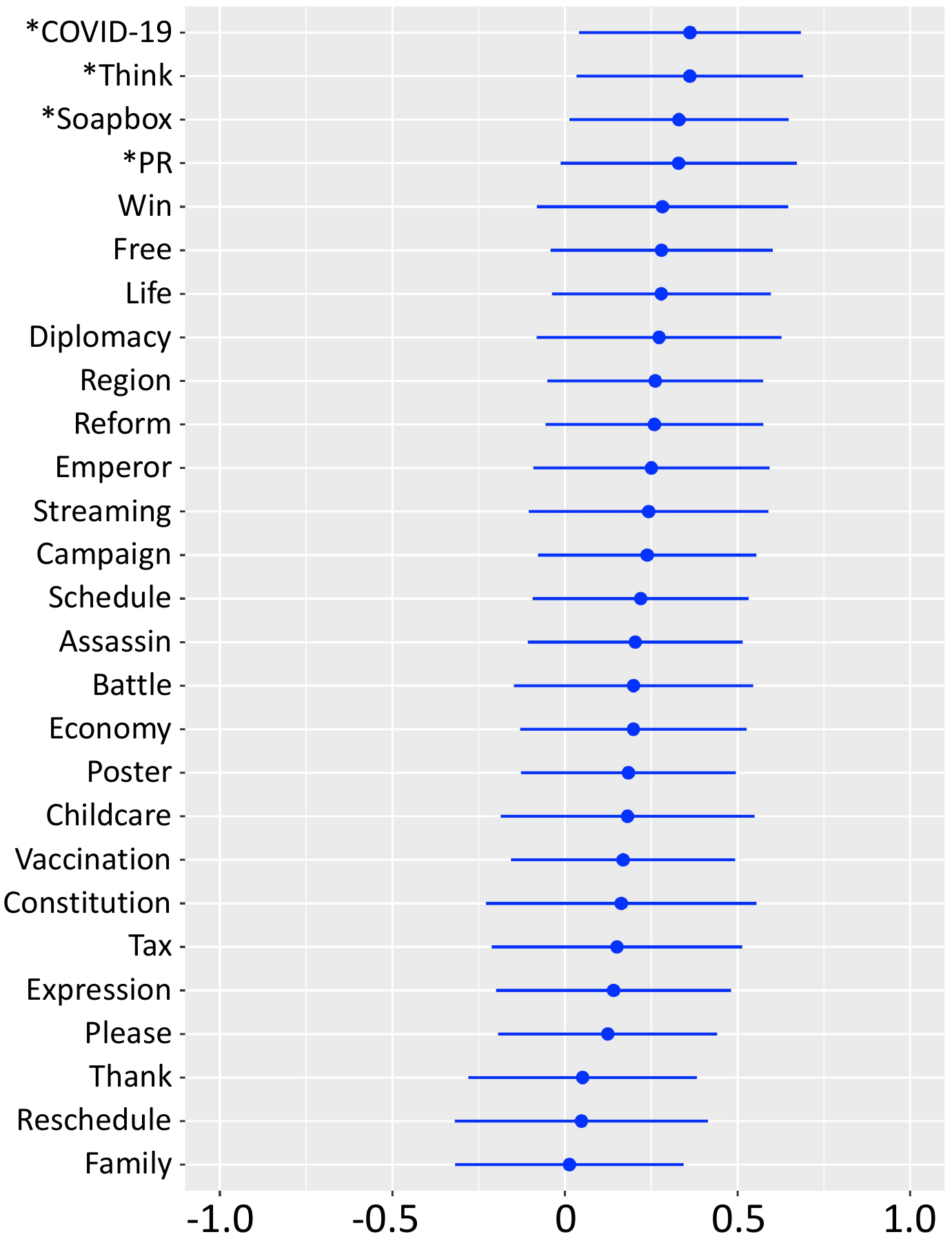}
    \subcaption{Group E in \textit{Election}}
  \end{subfigure}
  \hfill
  \begin{subfigure}[b]{0.24\linewidth}
    \centering
    \includegraphics[width=\textwidth]{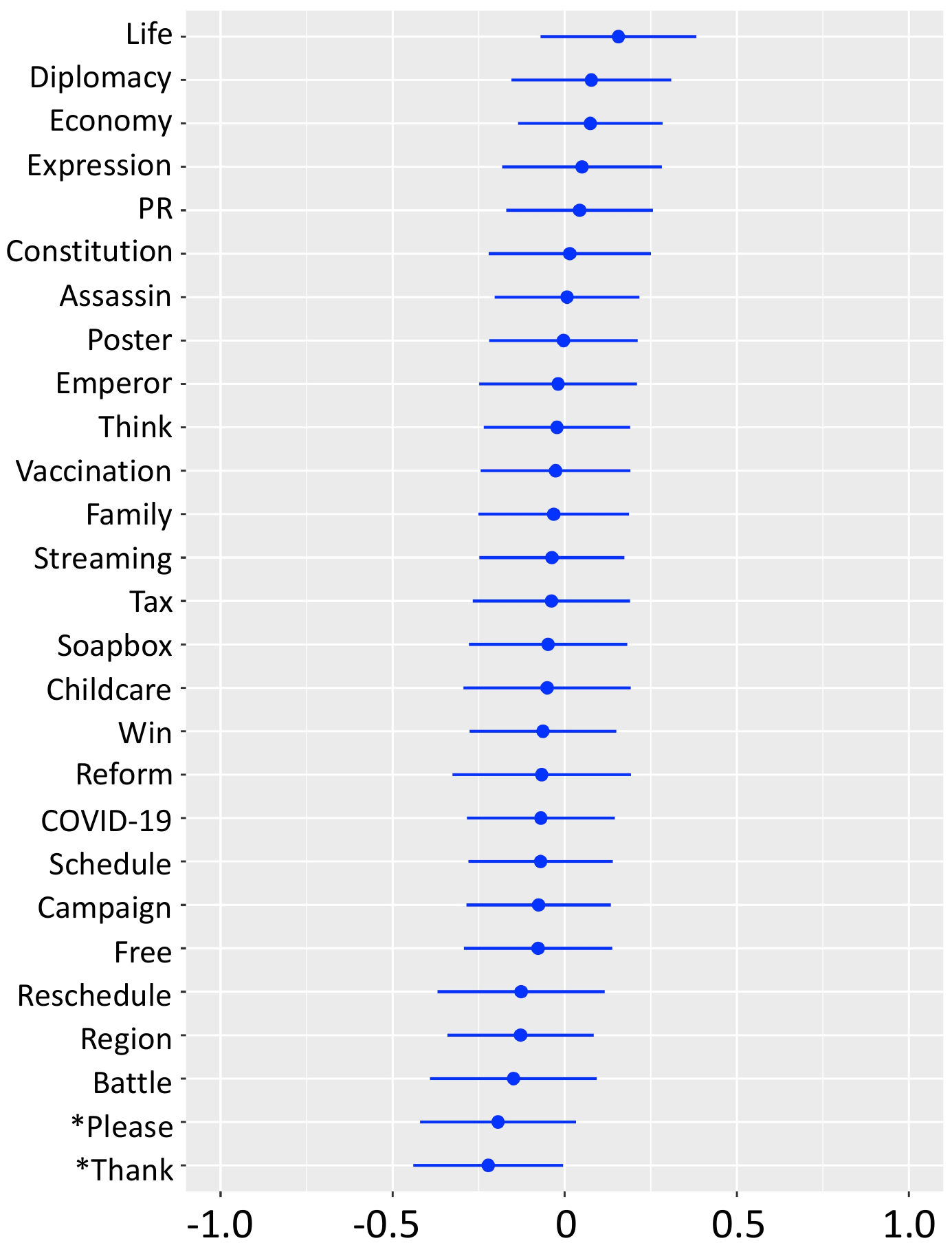}
    \subcaption{Group L in \textit{Election}}
  \end{subfigure}
  \hfill
  \begin{subfigure}[b]{0.24\linewidth}
    \centering
    \includegraphics[width=\textwidth]{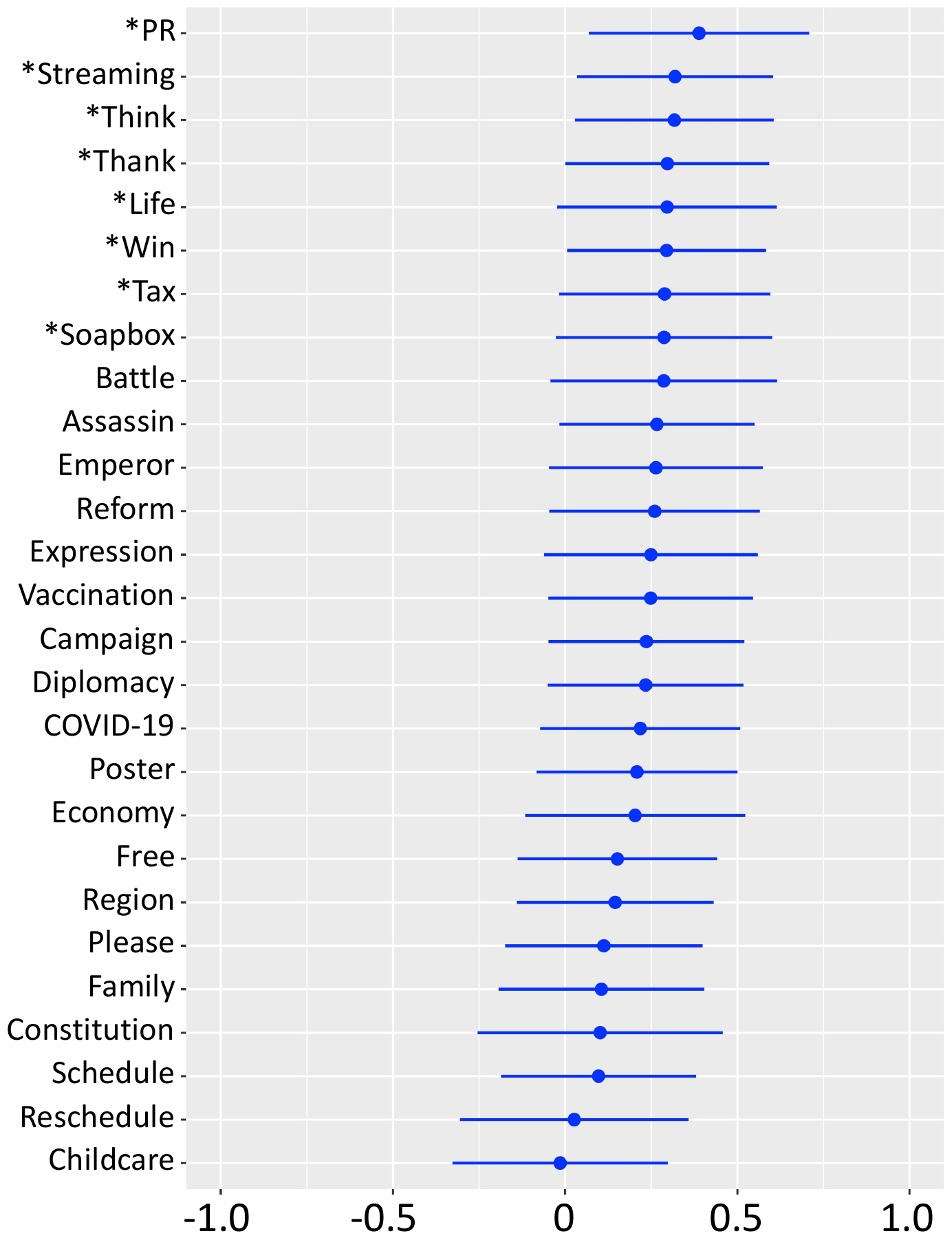}
    \subcaption{Group PR in \textit{Election}}
  \end{subfigure}
  
  \caption{All topics sorted by effect size ($\beta_{1}$) that predicts the number of user engagement (retweets) by Equation~\ref{equ} in each group and each term.
  }
  \label{regression2}
\end{figure*}

\subsection{Regression model}
What type of content is likely to gain user engagement in each group?
This section examines the trade-off between tweet topics and user engagement using a linear mixed-effects model~\cite{gelman2006data,bates2014fitting}.
The data is analyzed at the tweet level, while mixed effects account for variability across the characteristics of the user.
Concretely, we set the number of user engagements measured by the number of retweets\footnote{We also performed a regression analysis of the number of likes as an independent variable, but the results were similar to those of the number of retweets as an independent variable in Figure~\ref{regression2}. This section shows only the results for the number of retweets.} as the independent variable and apply a log transformation to reduce the influence of extremes.
We set the topic of the tweet, obtained in Section 5, as the explanatory dummy variable.
Moreover, we set a random intercept per each candidate because the numbers of followers are highly correlated with user engagements~\cite{uysal2011user}.
We use their political party as a control variable to mitigate the influence of the political party~\cite{keller2018followers,blassnig2021popularity}.
The regression model is defined as below;
\begin{align}
\label{equ}
    \log (Engagement + 1) = \beta_{1} * Topic  \nonumber \\
                + \beta_{2} * Party + \phi + \epsilon  
\end{align}
where $Engagement$ is the number of retweets, $\phi$ is the random effect for one of all the candidates, and $\epsilon$ is the error term.
We report the effect size $\beta_{1}$, which is the coefficients of all topics, for each group in each term by fitting the model to their tweets.

\subsection{Results}

We present all topics sorted by effect size in each group in Figure~\ref{regression2}.
During \textit{Pre-Election} term, topics regarding political policies tend to get more retweets; Tax, Expression (the topic on the regulation of expression), Economy, Childcare, and Emperor (the topic on the Emperor System).
On the Constitution topic that sparks national debate, tweets in group L and PR tend to get retweets, while those in group W are less likely to be retweeted.
The candidates in group W, who have already gained popularity with the public, may be less likely to make tweets that generate public interest on sensational topics for fear of social media ablaze.

During \textit{Election} term, topics regarding political policies tend to get more retweets, similar to the characteristics before the election announcement.
Just because it is an election period does not necessarily mean that election-related topics such as Campaign (the topic of campaign speech) and Schedule (the topic of reporting upcoming events) are likely to get a lot of retweets.
On the other hand, tweets in group E with no obvious election results tend to get more retweets for Soapbox (the topic of their own soapbox oratory) and PR (the topic of the support for proportional representation), suggesting that they are making an effort to spread information about the election.
Group W gets retweets, especially for tweets about Assassin, which is the topic of tweets about former Prime Minister Abe's assassination.
They are better than other groups at expressing sympathy and anger over sensational incidents and have gained support.
What is interesting is that Please, which is a distinctive topic in group L and PR (Table\ref{topic_kai}), is not a topic that can get retweets.
Candidates in these groups frequently ask their followers to vote and retweet, but it indicates that such tweets are unlikely to get a response from followers.

\section{RQ4: Is there a difference in the way they communicate with other users on social media?}
Figure~\ref{pre_election_tweet} and \ref{election_tweet} showed a significant difference in the level of activity in the use of the reply function in each group in Section 4.
This section attempts to analyze the replies that communicate directly with other users and political actors in more detail.

\begin{figure}[t]
  \begin{subfigure}[b]{0.48\linewidth}
    \centering
    \includegraphics[width=\textwidth]{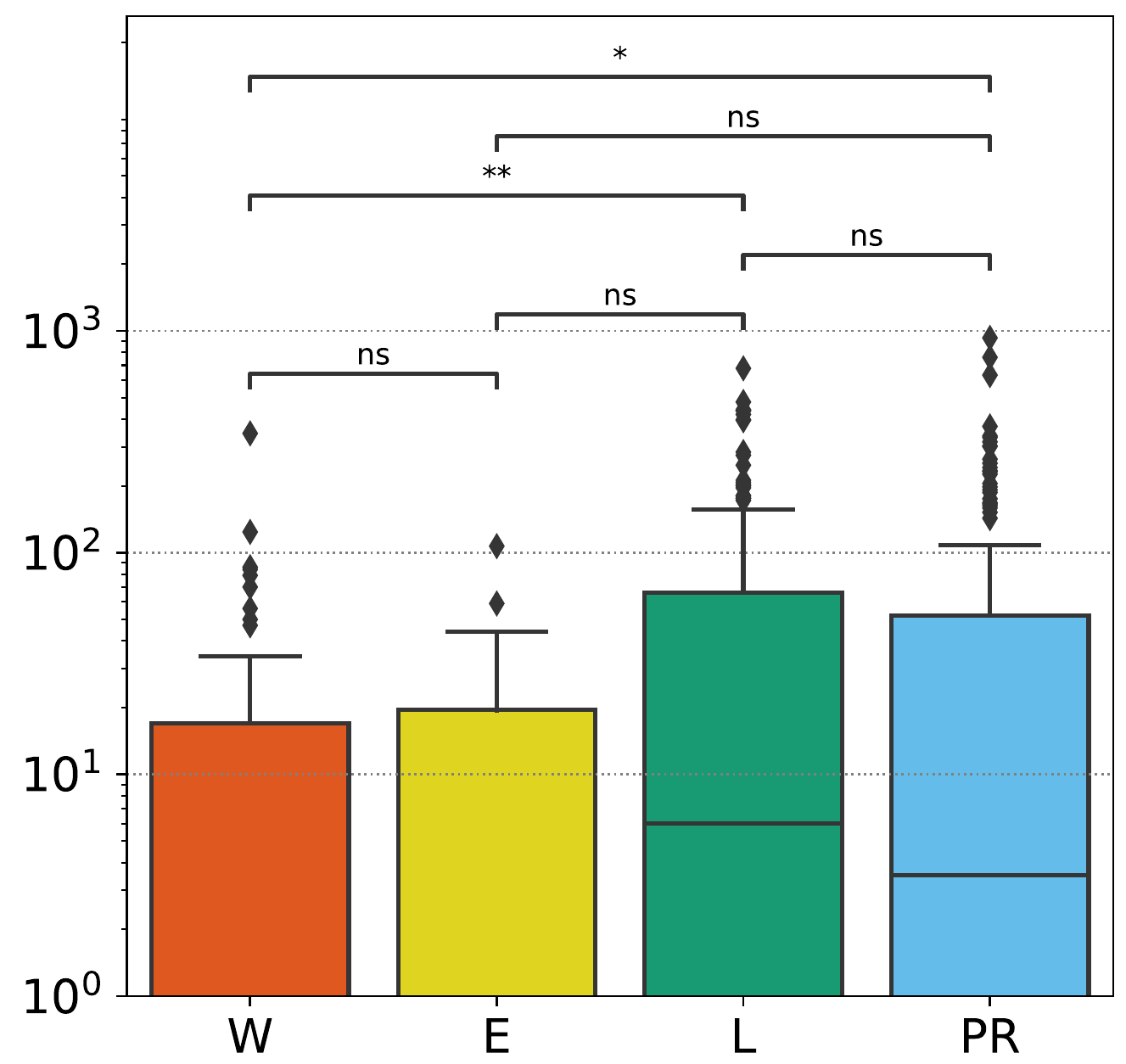}
    \subcaption{Number of reply target users during \textit{Pre-Election} term}
  \end{subfigure}
  \hfill
  \begin{subfigure}[b]{0.48\linewidth}
    \centering
    \includegraphics[width=\textwidth]{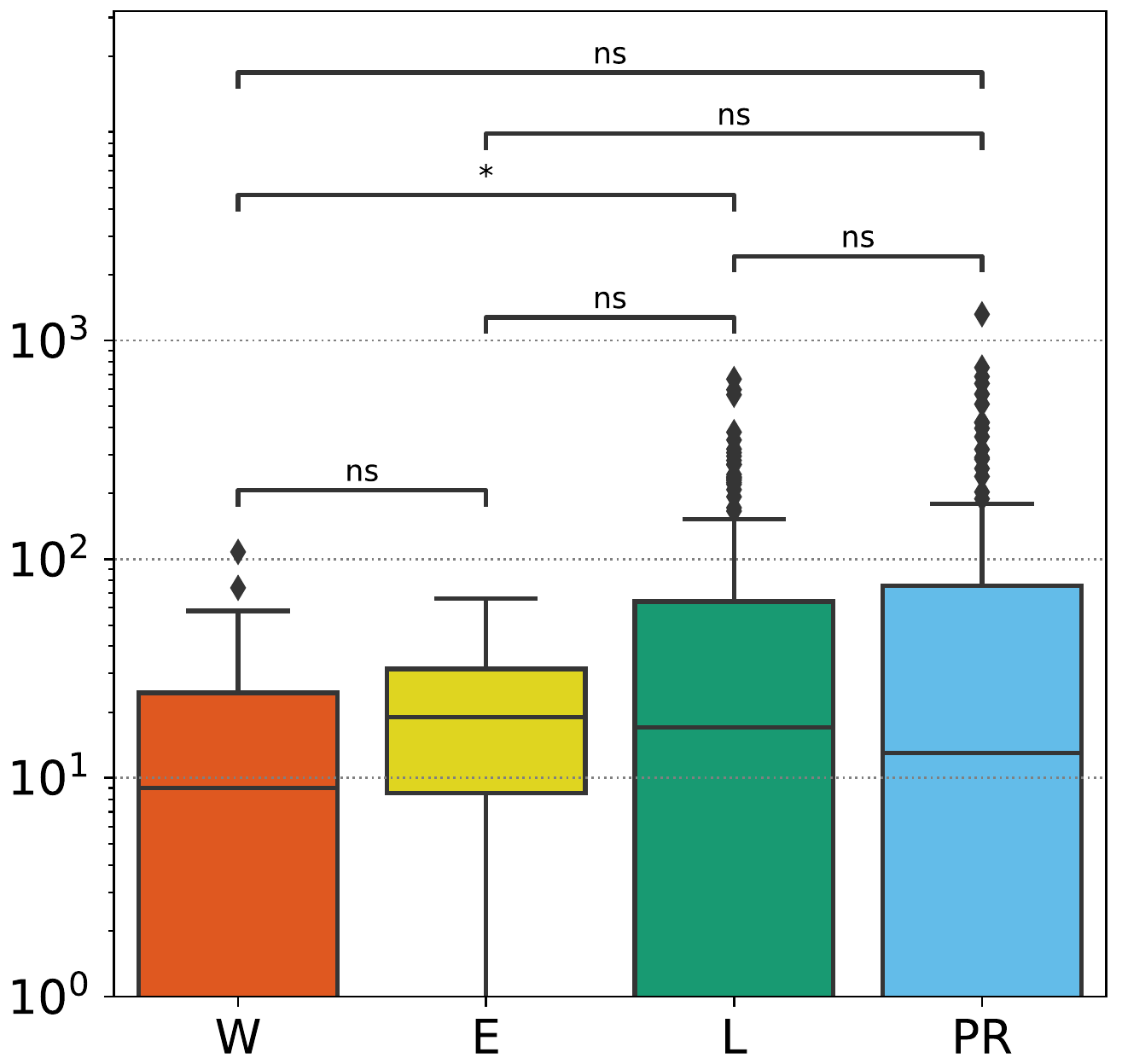}
    \subcaption{Number of reply target users during \textit{Election} term}
  \end{subfigure}
  \\
  \begin{subfigure}[b]{\linewidth}
    \centering
    \includegraphics[width=\textwidth]{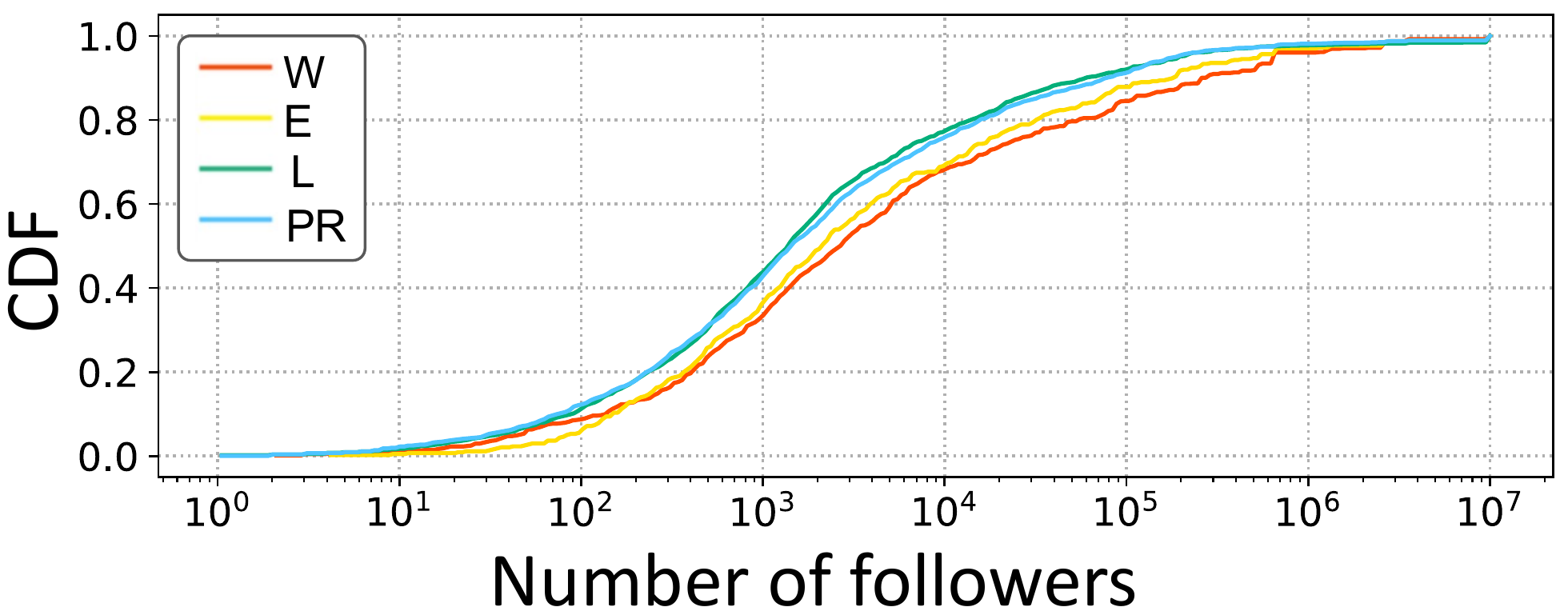}
     \subcaption{The cumulative distribution of number of followers in the reply target}
  \end{subfigure}
  \caption{Statistics on target users conversed with through each candidate's reply}
  \label{reply_target}
\end{figure}

\noindent \textbf{Reply target users}:
Figure~\ref{reply_target} (a) and (b) show the number of the users to whom candidates in each group replied during \textit{Pre-Election} and \textit{Election} terms.
Similar to the number of replies, groups W and E have far fewer users replied to than groups L and PR.
Groups E and W had fewer than 10 users conversing via the reply function during the election period, while some in Group L conversed with more than 10 users.
Then, figure~\ref{reply_target} (c) shows the cumulative distribution of the number of followers, representing popularity on social media, of reply target users in each group.
It represents the tendency of the conversation with users who have a large number of followers, in order of groups W, E, PR, and L.
Concretely, while 68.30\% of the users conversed with by candidates in group W have less than 10,000 followers, 75.31\% of the users conversed with by candidates in group L have less than 10,000 followers.
In addition, the verification accounts of the reply target users are 39.54\%, 33.00\%, 18.51\%, and 17.7\% in groups W, E, L, and PR, respectively.
The percentage of candidates who have had at least one conversation with another candidate among their reply target users is 70.00\%, 73.91\%, 78.77\%, and 72.58\% in groups W, E, L, and PR.
These results indicate that candidates in groups W and E, where there is above a certain chance of winning, focus more on conversations with users with a verified badge and with many followers than general users.
In other words, they are using social media for broadcasting, not for interacting with voters~\cite{graham2013between}.
The finding that group E, who is in a state of close competition, has fewer interactions with other users than groups L and PR differs from the finding in previous study~\cite{kahn2022spectacle}, that the more intense the election campaign, the stronger the interaction with voters.

\begin{figure}[t]
  \begin{subfigure}[b]{\linewidth}
    \centering
    \includegraphics[width=\textwidth]{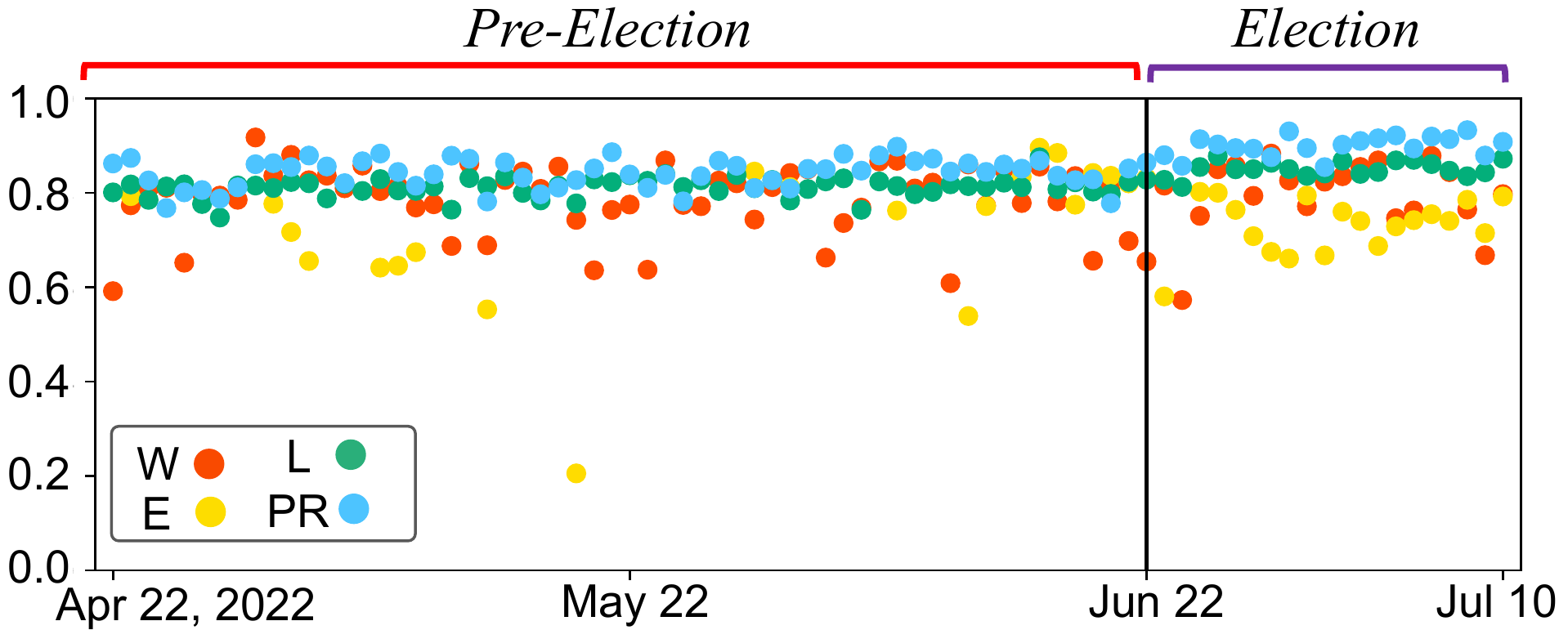}
    \subcaption{Time series of sentiment score}
  \end{subfigure}
  \\
  \begin{subfigure}[b]{\linewidth}
    \centering
    \includegraphics[width=\textwidth]{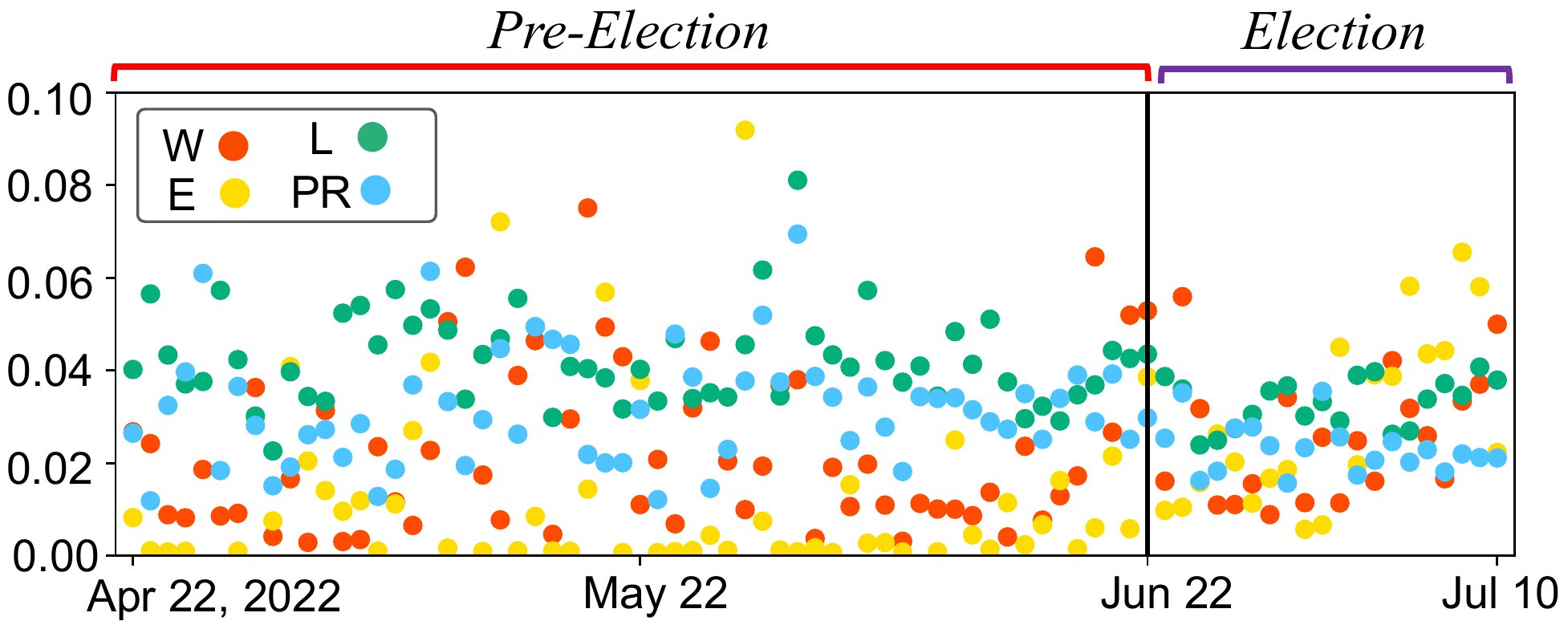}
    \subcaption{Time series of toxicity score}
  \end{subfigure}
  \caption{Time series of the average of (a) sentiment and (b) toxicity score of replies in each group. The average score per day is expressed as a single point.
  The point where there are less than 5 tweets in a day has been removed.}
  \label{time_reply}
\end{figure}

\noindent \textbf{Sentiment and toxicity score in reply tweets}
The form of presentation such as whether the reply is positive or negative also significantly impacts voters' impressions~\cite{ferrucci2020civic}.
We investigate each group's replies from two perspectives: sentiment and toxicity.
For the sentiment analysis, we use Asari~\cite{asari}, which returns a sentiment value between 0 and 1 (the closer to 1, the more positive) when a Japanese sentence is inputted.
Asari, which is an open-source sentiment quantification model based on SVM, is reported to perform with compelling accuracy as BERT-based models and is faster.
For the toxicity analysis, we use Jigsaw's Perspective API~\cite{perspective}, which is widely used for hate speech detection, to measure text toxicity using machine learning.

Figure~\ref{time_reply} (a) shows the time series of the average score of replies in sentiment.
There is a tendency for the average score of replies to increase for \textit{Election} term than for \textit{Pre-Election} term for all groups, e.g., the average score of replies in group W increased from 0.786 to 0.821.
It means that more candidates are making positive replies during the election period than usual.
The comparisons between groups show a high average score in the order of Groups PR, L, W, and E.
For example, during \textit{Election} term, the difference is remarkable, with an average score of 0.897 for group PR compared to 0.820 for Group E.
It is interesting to note that the candidates who are more likely to be elected have fewer positive replies than those to run in the proportional representation or with little chance of winning the election.

Figure~\ref{time_reply} (b) shows the time series of the average score of replies in toxicity.
The change in toxicity scores over the two terms tended to vary by the group; in groups W and E, the toxicity score increases from \textit{Pre-Election} term to \textit{Election} term ($0.022 \rightarrow 0.027$ and $0.019 \rightarrow 0.030$), while in groups L and PR, it decreases ($0.042 \rightarrow 0.034$ and $0.031 \rightarrow 0.023$).
The fact that there were more toxic replies in the two groups runs counter to the intuition to be cautious in replying to other users during the elections and requires further investigation.

\section{Limitation and future work}
This research attempts to shed light on the relationship between the chance of winning an election and candidates' political communication on social media.
However, it is necessary to acknowledge that there are several limitations to be considered.
The first limitation is that it is unclear whether our findings targeting candidates in the Japanese election can be generalized to those in other countries.
The application of our analysis to candidates' tweets in other countries may lead to new findings; therefore, the analysis of other countries with different electoral systems and social backgrounds is also expected to be conducted. 
Second, we used data for the three months prior to the election, but it is not clear whether similar conclusions can be drawn for a period far from the election period or for other events.
To clarify this point, it is essential to work with large datasets over a long period in the future.
Finally, there is still much room to analyze the tweet contents.
Our research focused on the topics, toxicity, and sentiment of their tweets; furthermore, by utilizing natural language processing techniques such as rhetorical analysis, stance detection, or discourse framing, we expect to grasp clues to understand the candidates’ intentions from the text in their tweets. \looseness=-1

\section{Conclusions}
Our study advances our understanding of how the chance of winning an election affects political communication on social media.
We grouped election candidates into four groups according to the chance and characterized their social media strategy in terms of users, topics, and reply behavior.
Our analysis showed that the attitude with which they engage in political communication and the topics they talk about differ depending on the chance of winning an election.
Furthermore, it discovered, as their chances of winning increase, candidates narrow the targets they communicate with, from people in general to the electoral districts and specific persons.
Our findings highlighting candidate behavior from a new perspective are helpful for future election strategies. \looseness=-1

\section*{Ethical Considerations}
The data in this paper is derived from publicly-accessible user-generated content.
We pay the utmost attention to the privacy of individuals in this study. 
When sharing our twitter data, we will publish only a list of tweet IDs. \looseness=-1

\section*{Acknowledgement}
This work was supported by
JSPS KAKENHI Grant-in-Aid for Scientific Research Number
JP20H00585,    
JP21H03446,    
JP22K20159, 
JP23K16889, 
NICT 03501, 
MIC/SCOPE JP192107004, 
JST-AIP JPMJCR21U4. 

\bibliography{ref}

\end{document}